\documentclass[10pt,letterpaper]{article}

\usepackage{subfigure}

\usepackage[top=0.85in,left=2.75in,footskip=0.75in]{geometry}

\usepackage{amsmath,amssymb}

\usepackage{changepage}

\usepackage[utf8x]{inputenc}

\usepackage{textcomp,marvosym}

\usepackage{cite}

\usepackage{nameref,hyperref}

\usepackage[right]{lineno}

\usepackage{microtype}
\DisableLigatures[f]{encoding = *, family = * }

\usepackage[table]{xcolor}

\usepackage{array}

\newcolumntype{+}{!{\vrule width 2pt}}

\newlength\savedwidth

\raggedright
\usepackage[aboveskip=1pt,labelfont=bf,labelsep=period,justification=raggedright,singlelinecheck=off]{caption}

\makeatletter
\renewcommand{\@biblabel}[1]{\quad#1.}
\makeatother

\usepackage{lastpage,fancyhdr,graphicx}
\usepackage{epstopdf}
\fancyheadoffset[L]{2.25in}
\fancyfootoffset[L]{2.25in}
\begin{document}
\vspace*{0.2in}

\begin{flushleft}
{\Large
\textbf\newline{Empirical Evaluation of Typical Sparse Fast Fourier Transform Algorithms} 
}
\newline
\\
Bin Li\textsuperscript{1},
Zhikang Jiang\textsuperscript{1},
Jie Chen\textsuperscript{1*},
\\
\bigskip
\textbf{1} School of Mechanical and Electrical Engineering and Automation, Shanghai University, Shanghai 200072, China
\bigskip

* jane.chen@shu.edu.cn

\end{flushleft}
\section*{Abstract}
Computing the Sparse Fast Fourier Transform(sFFT) of a $K$-sparse signal of size $N$ has emerged as a critical topic for a long time. The sFFT algorithms decrease the runtime and sampling complexity by taking advantage of the signal's inherent characteristics that a large number of signals are sparse in the frequency domain. More than ten sFFT algorithms have been proposed, which can be classified into many types according to filter, framework, method of location, method of estimation. In this paper, the technology of these algorithms is completely analyzed in theory. The performance of them is thoroughly tested and verified in practice. The theoretical analysis includes the following contents: five operations of signal, three methods of frequency bucketization, five methods of location, four methods of estimation, two problems caused by bucketization, three methods to solve these two problems, four algorithmic frameworks. All the above technologies and methods are introduced in detail and examples are given to illustrate the above research. After theoretical research, we make experiments for computing the signals of different SNR, $N$, $K$ by a standard testing platform and record the run time, percentage of the signal sampled and $L_0,L_1,L_2$ error with eight different sFFT algorithms. The result of experiments satisfies the inferences obtained in theory.

\section*{Introduction}
The Discrete Fourier Transform(DFT) is one of the most important and widely used techniques in mathematical computing. It is a basic tool commonly used to analyze the spectral representation in signal processing. Hence, people have been committed to the research of fast DFT algorithm for a long time. The most popular algorithm to compute DFT is the fast Fourier transform(FFT) invented by Cooley and Tukey, which can compute a signal of size $N$ in $O(N\text{log}N)$ time and use $O(N)$ samples. With the demand for low sampling ratio and big data computing, it motivates the new algorithms to replace the previous FFT algorithms that can compute DFT from a subset of the input data in sub-linear time. The new algorithms called sFFT algorithms can reconstruct the spectrum with high accuracy by using only $K$ most significant frequencies. In terms of its excellent performance and generally satisfied assumptions, the technology of sFFT was named one of the ten Breakthrough Technologies in MIT Technology Review in 2012.

As is shown in Table \ref{tab1}, more than ten sFFT algorithms have been proposed. Now we have to figure out how these different types of sFFT algorithms perform in theory and in practice. The performance includes runtime complexity, sampling complexity, robustness. The comparative study of the performance of these algorithms is helpful to apply them to suitable scenes and for reasonable objects. All the answers can be found in this paper.

\begin{table}[!ht] 
\begin{adjustwidth}{-1in}{0in} 
\caption{Description of sFFT algorithm characteristics}
\setlength{\tabcolsep}{4pt}
\begin{tabular}{|p{50pt}|p{50pt}|p{50pt}|p{50pt}|p{50pt}|p{50pt}|p{55pt}|p{30pt}|}
\hline
Algorithm& 
Filter& 
Location & 
Estimation & 
Framework& 
Length& 
Deterministic & 
Scene

 \\
\hline
AAFFT0.9 &
Dirichlet & 
Binary search & 
Frequency shift &
Iteration &
Power of two & 
No & 
General sparse
\\

SFFT-DT&
Spike train & 
Prony & 
Prony & 
One-shot &
Power of two & 
No & 
General sparse
\\

FFAST&
Spike train & 
Phase encoding& 
Energy &
Peeling&
Product of numbers& 
Yes& 
Exactly sparse
\\

R-FFAST&
Spike train & 
Prony& 
Prony&
Peeling&
Product of numbers& 
Yes& 
General sparse
\\

sFFT1.0 &
Flat& 
Statistics& 
Energy&
Voting &
Power of two& 
No& 
General sparse
\\

sFFT2.0&
Spike train, Flat& 
Statistics& 
Energy&
Voting &
Power of two& 
No& 
General sparse
\\

sFFT3.0&
Flat& 
Phase encoding& 
Energy&
Iteration &
Power of two& 
Yes& 
Exactly sparse
\\

sFFT4.0&
Flat& 
Multi-scale search& 
Energy&
Iteration &
Power of two& 
Yes& 
General sparse
\\

MPFFT&
Flat& 
Binary search& 
Energy&
Iteration &
Power of two& 
Yes& 
General sparse
\\
\hline
\end{tabular}
\label{tab1}
\end{adjustwidth}
\end{table}

The sFFT algorithms using the Dirichlet kernel filter bank is a randomized algorithm. The performance of the Ann Arbor fast Fourier transform(AAFFT0.5 \cite{bib1}) algorithm was later improved in the AAFFT0.9 \cite{bib2}, \cite{bib3} algorithm through the use of unequally-spaced FFTs and binary search technique for spectrum reconstruction. 

There are two frameworks for the sFFT algorithms using the spike train filter. The algorithms of the one-shot framework based on the compressed sensing solver are the so-called sFFT by downsampling in the time domain(sFFT-DT1.0 \cite{bib4}, sFFT-DT2.0 \cite{bib5}) algorithm. The algorithms of the peeling framework based on the bipartite graph are the so-called Fast Fourier Aliasing-based Sparse Transform(FFAST) \cite{bib6}, \cite{bib7} and R-FFAST \cite{bib8}, \cite{bib9} algorithm. Under the assumption of arbitrary sampling, the Gopher Fast Fourier Transform(GFFT) \cite{bib10}, \cite{bib11} algorithm and the Christlieb Lawlor Wang Sparse Fourier Transform(CLW-SFT) \cite{bib12}, \cite{bib13} algorithm are aliasing-based search deterministic algorithms guided by the Chinese Remainder Theorem(CRT). The DMSFT \cite{bib14}(generated from GFFT) algorithm and CLW-DSFT \cite{bib14}(generated from CLW-SFT) algorithm use the multiscale error-correcting method to cope with noise.

There are two frameworks for the sFFT algorithms using the flat filter. The algorithms of the voting framework are the sFFT1.0 \cite{bib15} and sFFT2.0 \cite{bib15} algorithm which can locate and estimate the $K$ largest coefficients in voing by multiple random bucketization. The algorithms of the iterative framework are the sFFT3.0 \cite{bib16} and sFFT4.0 \cite{bib17} algorithm and etc. The sFFT3.0 algorithm can locate the position by using only two filtered signals inspired by the frequency offset estimation in the exactly sparse case. The sFFT4.0 algorithm can locate the position block by block inspired by the multiscale frequency offset estimation in the general sparse case. The new robust algorithm, so-called the Matrix Pencil FFT(MPFFT) algorithm, was proposed based on the sFFT3.0 algorithm. The paper \cite{bib18} summarizes the two frameworks and five reconstruction methods of these five corresponding algorithms.

The paper \cite{bib19} summarizes a three-step approach in the stage of spectrum reconstruction and provides a standard testing platform to evaluate different sFFT algorithms. There are also some researches try to conquer the sFFT problem from other aspects: complexity \cite{bib20}, \cite{bib21}, performance \cite{bib22}, \cite{bib23}, software \cite{bib24}, \cite{bib25}, hardware \cite{bib26}, higher dimensions \cite{bib27}, \cite{bib28}, implementation \cite{bib29}, \cite{bib30} and special setting \cite{bib31}, \cite{bib32} perspectives.

	This paper is structured as follows. Section 2 provides a brief overview of some notation and basic signal processing that we will use in the sFFT. Section 3 introduces and analyzes the first stage of sFFT, frequency bucketization, including three kinds of bucketizations through different filters. Section 4 introduces and analyzes the second stage of sFFT, spectrum reconstruction, including five methods of location and four methods of estimation. In Section 5, two confusions frequency aliasing and spectrum leakage caused by bucketization are raised and solved afterward, then four frameworks of sFFT algorithms are summarized, and the theoretical performance of their corresponding algorithms is analyzed. In Section 6, we use a standard platform to compare all kinds of sFFT algorithms to verify the theoretical research. We do not cover the analysis of detailed proof of all processes (the reader is referred to the original papers for proofs and analysis). However, we present an empirical analysis of the performance of the algorithms in theory and in practice. 

\section*{Notation}
\subsection*{Problem Statement of sFFT}
	In this section, we initially present some notation and basic definitions of sFFT. 

The $N$-th root of unify is denoted by $\omega _{N}=e^{-2\pi \mathbf{i}/N}$. The DFT matrix of size $N$ is denoted by $\mathbf{F}_{N}\in \mathbb{C}^{N\times N}$ as follows:	
\begin{equation}
\mathbf{F}_{N}[j,k]=\frac{1}{N}\omega _{N}^{jk} \label{Eq1}
\end{equation}

The DFT of a vector $x\in \mathbb{C}^{N}$ (consider a signal of size $N$) is a vector $\hat{x}\in \mathbb{C}^{N}$ defined as follows:
\begin{equation}
\begin{split}
\hat{x}=\mathbf{F}_{N}x	\\
\hat{x}_{i}=\frac{1}{N} \sum_{j=0}^{N-1}x_{j}\omega _{N}^{ij}	
\end{split}	\label{Eq2}
\end{equation}

For $x_{-i}=x_{N-i}$, the convolution is defined as follows:
\begin{equation}
(x\ast y)_{i}=\sum_{j=0}^{N-1}x_{j}y_{i-j} \label{Eq3}
\end{equation}

For the coordinate-wise product $(xy)_{i}=x_{i}y_{i}$, the DFT of $xy$ is performed as follows: 
\begin{equation}
\widehat{xy}=\hat{x}\ast \hat{y} \label{Eq4}
\end{equation}

In the exactly sparse case, spectrum $\hat x$ is exactly $K$-sparse if it has exactly $K$ non-zero frequency coefficients while the remaining $N-K$ coefficients are zero. In the general sparse case, spectrum $\hat x$ is general $K$-sparse if it has $K$ significant frequency coefficients while the remaining $N-K$ coefficients are negligible. The problem statement of sFFT is to recover a $K$-sparse approximation $\hat {x'}$ by locating $K$ frequency positions $f_0, \dots, f_{K-1}$ and estimating $K$ largest frequency coefficients $\hat{x}_{f_{0}},\dots,\hat{x}_{f_{K-1}}$. The error of approximation $\hat {x'}$  in approximating $\hat x$ is bounded by the error on the best $K$-sparse approximation $\hat{y}$ ($\hat{y}= \text{argmin}\left \| \hat{x} - \hat{y} \right \|_2$), Formally, $\hat {x'}$ satisfy the following $\ell_2/\ell_2$ guarantee where $C$ is an approximation factor:
\begin{equation}
\left \| \hat{x} - \hat {x'}\right \|_2\leq C\left \| \hat{x} - \hat{y}\right \|_2 \label{Eq5}
\end{equation}

\subsection*{Operation of signal}
As we all know, the preprocessing of sFFT requires multiple signal processing, including time shift operation, time scaling operation, frequency shift operation, subsampling operation, aliasing operation. These operations are equivalent to multiplying the original signal by a specified matrix. The details can be seen as follows:
\subsubsection*{Time shift operation}
The matrix representing the time shift operation is denoted by $\mathbf{S}_{\tau}\in \mathbb{R}^{N\times N} $ as follows:
\begin{equation}
\mathbf{S}_{\tau}[j,k]=\left\{\begin{matrix}  1, & j-\tau\equiv k(\text{mod} N) 
\\0,&\text{o.w.}
\end{matrix}\right. \label{Eq6}
\end{equation}

The time offset parameter is denoted by $\tau \in \mathbb{R}$. If a vector $x'=\mathbf{S}_{\tau} x$, such that $x'_i=x_{(i-\tau)}$.

\subsubsection*{Time scaling operation}
The matrix representing the time scaling operation is denoted by $\mathbf{P}_{\sigma}\in \mathbb{R}^{N\times N} $ as follows:
\begin{equation}
\mathbf{P}_{\sigma}[j,k]=\left\{\begin{matrix}  1, &  \sigma {j} \equiv k(\text{mod} N)
\\0,&\text{o.w.} 
\end{matrix}\right. \label{Eq7}
\end{equation}

The scaling parameter is denoted by $\sigma \in \mathbb{R}$. If a vector $x'=\mathbf{P}_{\sigma} x$, such that $x'_i=x_{\sigma i}$.

\subsubsection*{Frequency shift operation}
The matrix representing the frequency shift operation is denoted by $\mathbf{V}_{b}\in \mathbb{R}^{N\times N} $ as follows:
\begin{equation}
\mathbf{V}_{b}[j,k]=\left\{\begin{matrix}  \omega _{N}^{bj}, &  j=k
\\0,&\text{o.w.} 
\end{matrix}\right. \label{Eq8}
\end{equation}

The frequency offset parameter is denoted by $b \in \mathbb{R}$. If a vector $x'=\mathbf{V}_{b} x$, such that $x'_i=x_{ i}\omega _{N}^{bi}$.

\subsubsection*{Example and character of the above three operations}
Let us give an example to illustrate the above three operations. If a vector $x'=\mathbf{V}_{b} \mathbf{S}_{\tau} \mathbf{P}_{\sigma}x $, such that $x'_i=x_{\sigma(i-\tau)}\omega _{N}^{b' \sigma i}$ and $b= b' \sigma$. An example of the operations with three parameters $\tau =1, \sigma = 3, b' = 2, b= b' \sigma = 6$ is as follows:
\begin{tiny}
\begin{equation*}
\begin{split}
&\begin{bmatrix}
x'_0
\\ x'_1
\\x'_2
\\ x'_3
\\ \dots
\\ x'_{N-1}
\end{bmatrix}
=
\begin{bmatrix}
\omega _{N}^{0} & 0 & 0 & \dots & 0
\\ 0 & \omega _{N}^{6} & 0 & \dots & 0
\\0 & 0 & \omega _{N}^{12} & \dots & 0
\\ 0 & 0 & 0 & \dots & 0
\\ . & . & . & \dots & .
\\ 0 & 0 & 0 & \dots & \omega _{N}^{6(N-1)}
\end{bmatrix}
\begin{bmatrix}
0 & 0 & 0 & \dots & 1
\\ 1 & 0 & 0 & \dots & 0
\\0 & 1 & 0 & \dots & 0
\\ 0 & 0 & 1 & \dots & 0
\\ . & . & . & \dots & .
\\ 0 & 0 & 0 & \dots & 0
\end{bmatrix}
\begin{bmatrix}
1 & 0 & 0 &  0 & 0 & 0 & 0 & 0 & \dots & 0
\\ 0 & 0 & 0 & 1 & 0 & 0 & 0 & 0 &\dots & 0
\\0 & 0 & 0 & 0 & 0 & 0 & 0 & 1 &\dots & 0
\\ 0 & 0 & 0 & 0 & 0 & 0 & 0 & 0 &\dots & 0
\\ . & . & . & . & . & . & . & . &\dots & 0
\\ 0 & 0 & 0 & 0 & 0 & 0 & 0 & 0 &\dots & 0
\end{bmatrix}
\begin{bmatrix}
x_0
\\ x_1
\\x_2
\\ x_3
\\ .
\\ x_{N-1}
\end{bmatrix} \\
&\begin{bmatrix}
x'_0
\\ x'_1
\\x'_2
\\ x'_3
\\ \dots
\\ x'_{N-1}
\end{bmatrix}
=
\begin{bmatrix}
\omega _{N}^{0} & 0 & 0 & \dots & 0
\\ 0 & \omega _{N}^{6} & 0 & \dots & 0
\\0 & 0 & \omega _{N}^{12} & \dots & 0
\\ 0 & 0 & 0 & \dots & 0
\\ . & . & . & \dots & .
\\ 0 & 0 & 0 & \dots & \omega _{N}^{6(N-1)}
\end{bmatrix}
\begin{bmatrix}
0 & 0 & 0 &  0 & 0 & 0 & 0 & \dots  & 0
\\ 1 & 0 & 0 & 0 & 0 & 0 & 0 & \dots  & 0
\\0 & 0 & 0 & 1 & 0 & 0 & 0 & \dots  & 0
\\ 0 & 0 & 0 & 0 & 0 & 0 & 1 & \dots & 0
\\ . & . & . & . & . & . & . & \dots & .
\\ . & . & . & . & . & . & . & \dots  & .
\end{bmatrix}
\begin{bmatrix}
x_0
\\ x_1
\\x_2
\\ x_3
\\ .
\\ x_{N-1}
\end{bmatrix}\\
&\begin{bmatrix}
x'_0
\\ x'_1
\\x'_2
\\ x'_3
\\ \dots
\\ x'_{N-1}
\end{bmatrix}
=
\begin{bmatrix}
0 & 0 & 0 &  0 & 0 & 0 & 0 & \dots  & 0
\\ \omega _{N}^{6} & 0 & 0 & 0 & 0 & 0 & 0 & \dots  & 0
\\0 & 0 & 0 & \omega _{N}^{12} & 0 & 0 & 0 & \dots  & 0
\\ 0 & 0 & 0 & 0 & 0 & 0 & \omega _{N}^{18} & \dots & 0
\\ . & . & . & . & . & . & . & \dots & .
\\ . & . & . & . & . & . & . & \dots  & .
\end{bmatrix}
\begin{bmatrix}
x_0
\\ x_1
\\x_2
\\ x_3
\\ .
\\ x_{N-1}
\end{bmatrix}
=
\begin{bmatrix}
.
\\ x_0 \omega _{N}^{6}
\\x_3 \omega _{N}^{12}
\\ x_6 \omega _{N}^{18}
\\ .
\\ .
\end{bmatrix}
\end{split}
\end{equation*}
\end{tiny}

According to the property, the DFT of a random permutation signal is performed as follows: if $x'_i=x_{\sigma(i-\tau)}\omega _{N}^{b' \sigma i}$, such that we can get Equation (\ref{Eq9}). The proof of this equation is the same as the proof of Claim 3.5. in paper \cite{bib15}.
\begin{equation}
\hat{x'}_{\sigma( i-b') \text{mod}N}=\hat{x}_{i}\omega ^{\sigma \tau i} \label{Eq9}
\end{equation}

\subsubsection*{Subsampling operation in the time domain}
The matrix representing the subsampling operation is denoted by $\mathbf{D}_{L}\in \mathbb{R}^{B\times N}$ as follows:
\begin{equation}
\mathbf{D}_{L}[j,k]=\left\{\begin{matrix}  1, & k=jL
\\0,&\text{o.w.}
\end{matrix}\right. \label{Eq10}
\end{equation}

The subsampling parameter is denoted by $L\in\mathbb{Z}^{+}$, and the subsampling number is denoted by $B\in\mathbb{Z}^{+}$. If a vector $x'=\mathbf{D}_{L} x$, such that $x'_i=x_{ Li}$ and $\hat{x'}_{i} =  \hat{x}_{i}+\hat{x}_{i+B}+ \dots +\hat{x}_{i+(L-1)B}$. It is proved that the signal in the time domain is subsampled such that the corresponding signal in the frequency domain is aliased. In addition, this operation can also be seen as a signal multiplied by a spike train filter just shown in Fig \ref{fig3}(b). 

\subsubsection*{Aliasing operation in the time domain}
The matrix representing the aliasing operation is denoted by $\mathbf{U}_{L}\in \mathbb{R}^{B\times N}$ as follows:
\begin{equation}
\mathbf{U}_{L}[j,k]=\left\{\begin{matrix}  1, & j-k\equiv 0(\text{mod} B) 
\\0,&\text{o.w.}
\end{matrix}\right. \label{Eq11}
\end{equation}

The aliasing parameter is denoted by $L\in\mathbb{Z}^{+}$, and the aliasing number is denoted by $B\in\mathbb{Z}^{+}$. If a vector $x'=\mathbf{U}_{L} x$, such that $x'_{i} =x_{i}+x_{i+B}+ \dots + x_{i+(L-1)B}$ and $\hat{x'}_{i} =  \hat{x}_{iL}$. It is proved that the signal in the time domain is aliased such that the corresponding signal in the frequency domain is subsampled. In addition, this operation can also be regarded as the signal convoluted with a spike train filter.

\section*{The first Stage of sFFT: frequency bucketization}
The first stage of sFFT is encoding by frequency bucketization. In order to decrease runtime and sampling complexity, almost all operations are calculated in low dimensions through bucketization. Take advantage of this idea, sFFT algorithms work by binning $N$ Fourier coefficients into $B$ buckets($B \ll N$) at first. It is called bucketization. Since the signal is sparse in the frequency domain, one bucket has none significant coefficient in most cases and has only one significant coefficient in a small number of cases and has more than one significant coefficients only in very little cases, so we only have to deal with a little of effective buckets. The subsequent calculation in these buckets is mainly composed of two parts: location(to find its position) and estimation(to compute its value).

In order to save the run time and sampling number in frequency bucketization, the window filter used is required concentrated both in time and frequency domain. There are three types of filters that meet the requirements.

\subsection*{Introduction of flat filter}
\begin{figure} [!ht]\centering  
\subfigure[The Gaussian filter.]{  \includegraphics[width=0.31\columnwidth]{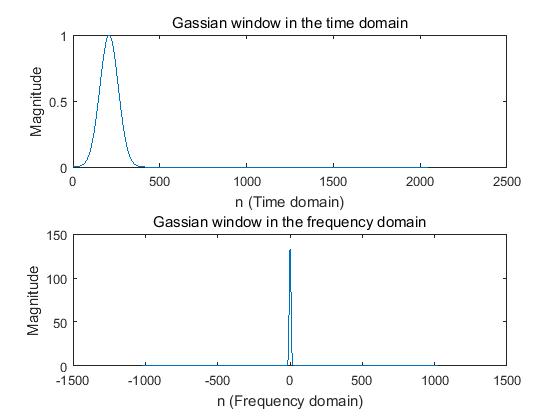} } \subfigure[The sinc filter.] { \includegraphics[width=0.31\columnwidth]{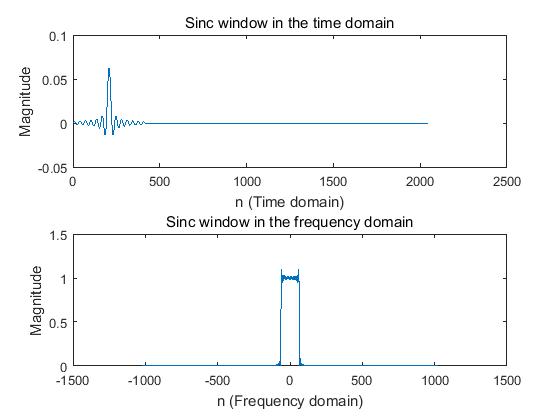} } \subfigure[The flat filter.] {\includegraphics[width=0.31\columnwidth]{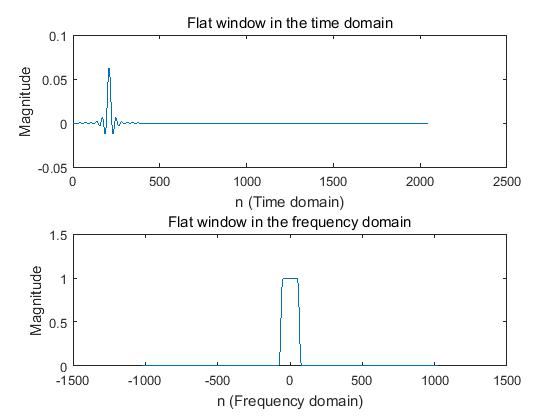} }
\caption{Three examples of a Gaussian filter, a sinc filter, a flat filter in the time domain and frequency domain.}  \label{fig1} 
\end{figure}

The flat filter looks like a box or rectangle function in the frequency domain while using a small number of samples in the time domain. An example of such filter is a sinc function multiplied by a Gaussian function in the time domain. The expressions of sinc function and Gaussian function are described as Equation (\ref{Eq12}), Equation (\ref{Eq13}). The examples of Gaussian, sinc, flat filter in the time domain and frequency domain with parameters $N=2048, L=128, B=16, \sigma = B\sqrt{\text{log}_2N} \approx  53.07 , \Omega  = 516$  are shown in Fig \ref{fig1}. From Equation (\ref{Eq4}), it is known that the product in the time domain is equal to convolution in the frequency domain, so as shown in Fig \ref{fig1}, the flat filter has almost no spectrum leakage. 
\begin{equation}
g_n=\left\{\begin{matrix}  1/B, & n=\frac{\Omega}{2}
\\ \frac{\sin(\pi /(B(n-\frac{\Omega}{2})))}{\pi(n-\frac{\Omega}{2})}, & n\in [0,\Omega -1]\cup n\neq \frac{\Omega}{2}
\\0,&\text{o.w.}
\end{matrix}\right. \label{Eq12}
\end{equation}
\begin{equation}
g_n=\left\{\begin{matrix}  e^{-\frac{1}{2}((n-\frac{\Omega}{2})/\sigma)^2}, & n\in [0,\Omega -1]
\\0,&\text{o.w.}
\end{matrix}\right. \label{Eq13}
\end{equation}

\subsection*{The analysis of bucketization through the flat filter}
\begin{figure} [!ht]\centering  
\subfigure[The original signal.]{  \includegraphics[width=0.45\columnwidth]{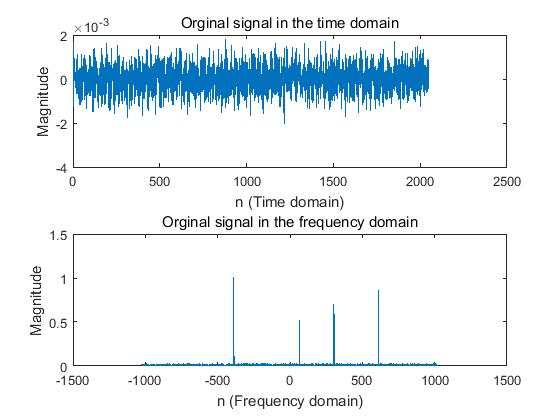} } \subfigure[The original signal and filtered signal.] { \includegraphics[width=0.45\columnwidth]{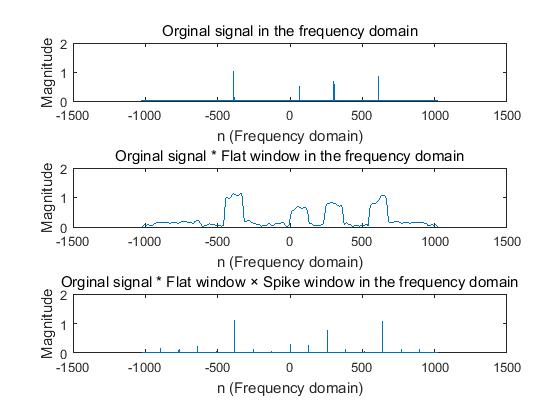} } 
\caption{The original signal and filtered signal through a flat filter in the frequency domain.}  \label{fig2} 
\end{figure}

In order to explain the techniques used in the bucketization through flat filter clearly, we start with the analysis with a typical application of an example. The original signal of the following experiments is considered as a $N(N = 2048)$ size signal that has only $K(K = 4)$ significant coefficients $\hat x _{64}=0.55, \hat x _{304}=0.7, \hat x _{610}=0.85, \hat x _{1660}=1.0$ while the rest of the coefficients are white Gaussian noise approximately equal to zero. The original signal in the time domain and frequency domain is shown in Fig \ref{fig2}(a). The filtered signal through a flat filter is shown in Fig \ref{fig2}(b).

As is shown in Fig \ref{fig2}(b), the filtered signal is obtained by multiplying the original signal with a flat filter and convoluted with a spike train filter in the time domain. It can also be obtained by convoluting the original signal with a flat filter and multiplying it with a spike train filter in the frequency domain. The filtered signal in $B$ dimension can be regarded as $B$ buckets. In each bucket, the filtered spectrum is mainly composed of $L$ neighboring elements. For example, big value $\hat x _{64},\hat x _{304},\hat x _{610},\hat x _{1660}$ is divided into bucket 1 (1=round(64/128)), bucket 2 (2=round(304/128), bucket 5 (5=round(610/128)), bucket 13 (13=round(1660/128)) individually. (round() means to make decimals rounded). The Equation (\ref{Eq14}) respecting the composition of filtered signal in bucket i can be obtained. The proof of this equation is same as the proof of Lemma 1 in paper \cite{bib18} (In the figure, $\ast$ represents the convolution of the signal and the filter in the frequency domain, $\times$ represents the multiplication of the signal and the filter in the frequency domain).
\begin{equation}
\hat{y}_{L, \tau, \sigma}[i] \approx \hat{G}_{\frac{L}{2}} \hat{x}_{\sigma^{-1}\left(\frac{(2i-1)L)}{2}\right)} \omega_{N}^{\tau\left(\frac{(2i-1)L}{2}\right)}+\ldots+\hat{G}_{-\frac{L}{2}+1} \hat{x}_{\sigma^{-1}(\frac{(2i+1)L}{2}-1)} \omega_{N}^{\tau (\frac{(2i+1)L}{2}-1)} \label{Eq14}
\end{equation}

\subsection*{Introduction of spike train filter}
\begin{figure} [!ht]\centering  
\subfigure[The spike train filter.]{  \includegraphics[width=0.45\columnwidth]{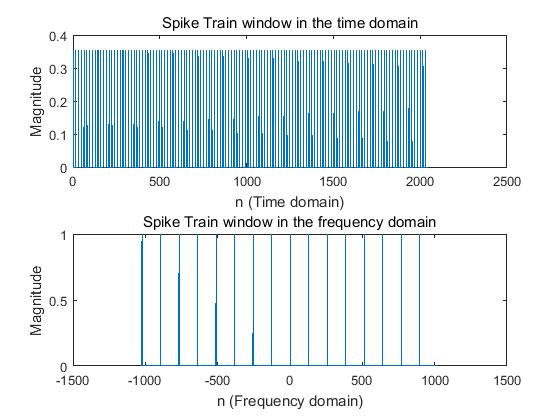} } \subfigure[The original signal and filtered signal.] { \includegraphics[width=0.45\columnwidth]{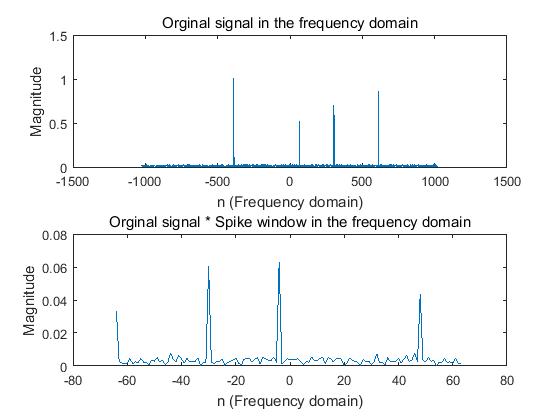} } 
\caption{The original signal and filtered signal through a spike train filter in the frequency domain.}  \label{fig3} 
\end{figure}

The spike train filter looks like uniformly distributed impulses in the time domain and frequency domain. The expression of spike train function is described as Equation (\ref{Eq15}). The example of a spike train filter in the time domain and frequency domain with parameter $N=2048, L=16, B=128$ is shown in Fig \ref{fig3}(a). 
\begin{equation}
g_n=\left\{\begin{matrix}  \sqrt{N}/L, & n\text{mod}B=0
\\0,&\text{o.w.}
\end{matrix}\right. \label{Eq15}
\end{equation}

\subsection*{The analysis of bucketization through spike train filter}
We start with the analysis with a typical application of an example. The filtered signal through a spike train filter is shown in Fig \ref{fig3}(b).

There are two applications of spike train filter. In one case, the filtered signal is equal to the product of the original signal and the spike train filter in the time domain, it can be represented by $x'=\mathbf{D}_{L} x$ and it is equivalent to aliasing in the frequency domain. In the other case, the filtered signal is equal to the convolution of the original signal and the spike train filter in the time domain, it can be represented by $x'=\mathbf{U}_{L} x$ and it is equivalent to subsampled in the frequency domain. Generally, bucketization through the spike train filter uses the first method, so the filtered signal is obtained by multiplying the original signal with a spike train filter in the time domain. In each bucket, the filtered spectrum is mainly composed of $L$ elements with the same remainder divided by $B$. The Equation (\ref{Eq16}) respecting the composition of filtered signal in bucket $i$ can be obtained. As is shown in Fig \ref{fig3}(b), big value $\hat x _{64},\hat x _{304},\hat x _{610},\hat x _{1660}$ is divided into bucket 64 (64=64mod128), bucket 48 (48=304mod128), bucket 98 (98=610mod128), bucket 124 (124=1660mod128) individually. 
\begin{equation}
\hat y _{B,\tau }[i]=\sum_{j=0}^{L-1}{\hat {x} _{jB+i} \omega ^{\tau(jB+i)}} \label{Eq16}
\end{equation}

\subsection*{Introduction of Dirichlet kernel filter bank}
\begin{figure} [!ht]\centering  
\subfigure[The Dirichlet kernel filter.]{  \includegraphics[width=0.45\columnwidth]{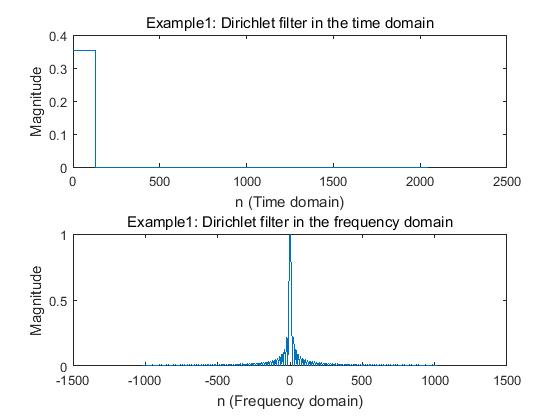} } \subfigure[The Dirichlet kernel filter.] { \includegraphics[width=0.45\columnwidth]{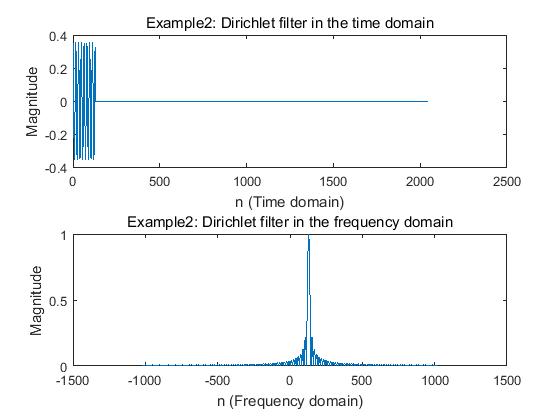} } 
\caption{Two examples of two Dirichlet kernel filters in the time domain and frequency domain.}  \label{fig4} 
\end{figure}

\begin{figure} [!ht]\centering  
\subfigure[The Dirichlet kernel filter bank.]{  \includegraphics[width=0.45\columnwidth]{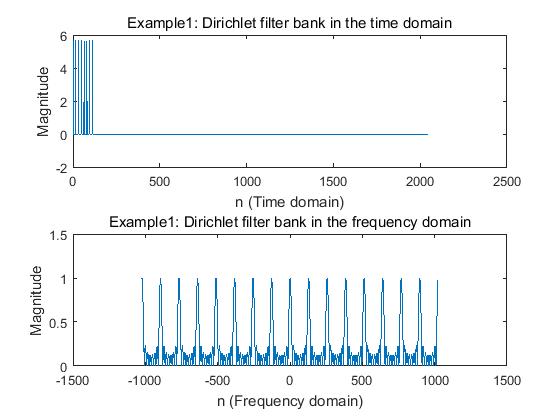} } \subfigure[The Dirichlet kernel filter bank.] { \includegraphics[width=0.45\columnwidth]{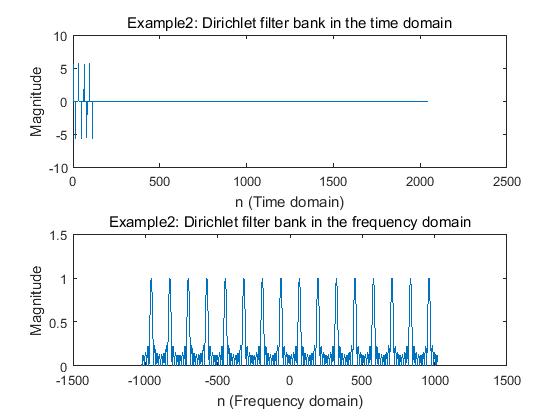} } 
\caption{Two examples of two Dirichlet kernel filter banks in the time domain and frequency domain.}  \label{fig5} 
\end{figure}

The Dirichlet kernel filter looks like a box or rectangle function in the time domain. The Dirichlet kernel filter bank is composed of K different Dirichlet kernel filter. Two examples of Dirichlet kernel function and Dirichlet kernel filter bank function are described as Equation (\ref{Eq17}), Equation (\ref{Eq18}), Equation (\ref{Eq19}), Equation (\ref{Eq20}). The examples of Dirichlet kernel filter and Dirichlet kernel filter bank in the time domain and frequency domain with parameter $N=2048, L=128, B=16$ are shown in Fig \ref{fig4} and Fig \ref{fig5}.
\begin{equation}
g_n=\left\{\begin{matrix}  \sqrt{N}/L, & n\in [0,L -1]
\\0,&\text{o.w.}
\end{matrix}\right. \label{Eq17}
\end{equation}
\begin{equation}
g_n=\left\{\begin{matrix}  (\sqrt{N}/L) \omega _N^n, & n\in [0,L -1]
\\0,&\text{o.w.}
\end{matrix}\right. \label{Eq18}
\end{equation}
\begin{equation}
g_n=\left\{\begin{matrix}  (\sqrt{N}/L) (\omega _N^{0}+\omega _N^n+\omega _N^{2n}+\dots+\omega _N^{(B-1)n}), & n\in [0,L -1]
\\0,&\text{o.w.}
\end{matrix}\right. \label{Eq19}
\end{equation}
\begin{equation}
g_n=\left\{\begin{matrix}  (\sqrt{N}/L) (\omega _N^{0.5n}+\omega _N^{1.5n}+\omega _N^{2.5n}+\dots+\omega _N^{(B-0.5)n}), & n\in [0,L -1]
\\0,&\text{o.w.}
\end{matrix}\right. \label{Eq20}
\end{equation}

\subsection*{The analysis of bucketization through Dirichlet kernel filter bank}
\begin{figure} [!ht]\centering  
\subfigure[The original signal and filtered signal.]{  \includegraphics[width=0.45\columnwidth]{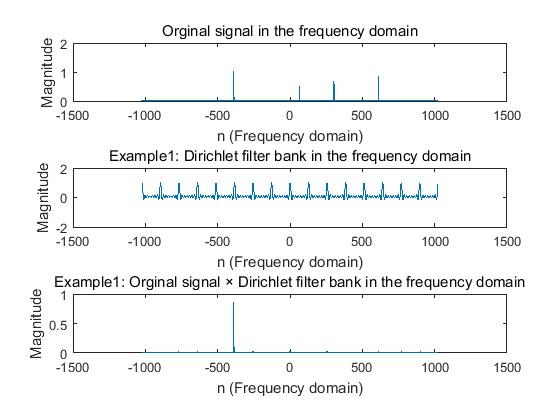} } \subfigure[The original signal and filtered signal.] { \includegraphics[width=0.45\columnwidth]{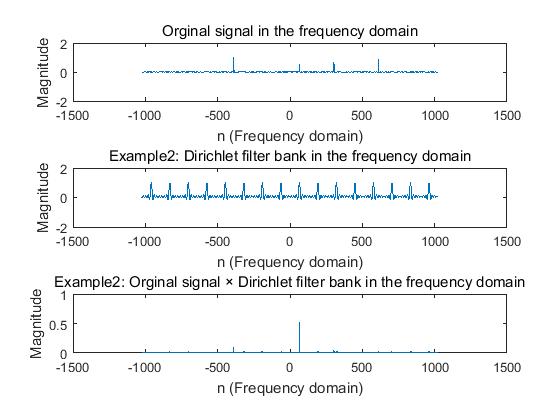} } 
\caption{Two examples of the original signal and filtered signal through Dirichlet kernel filter bank in the frequency domain.}  \label{fig6} 
\end{figure}

We start with the analysis with a typical application of an example. The filtered signal through a Dirichlet kernel filter bank is shown in Fig \ref{fig6}.

As is shown in Fig \ref{fig6}, the filtered signal is obtained by the convolution of the original signal and the Dirichlet kernel filter bank in the time domain. It can also be obtained by the product of the original signal and the Dirichlet kernel filter bank in the frequency domain. In each bucket, the filtered spectrum is mainly composed of $L$ neighboring elements. As is shown in Fig \ref{fig6}(a), big value $\hat x _{64},\hat x _{304},\hat x _{610},\hat x _{1660}$ is divided into bucket 1 (1=Int((64+64)/128)), bucket 2 (2=Int((304+64)/128)), bucket 5 (5=Int((610+64)/128)), bucket 13 (13=Int((1660+64)/128))) individually (Int() means to get only integer part). As is shown in Fig \ref{fig6}(b), big value is divided into bucket 0 (0=Int((64+0)/128), bucket 2 (2=Int((304+0)/128)), bucket 4 (4=Int((610+0)/128), bucket 12 (12=Int((1660+0)/128))) individually. In the energy of a bucket, the proportion coefficients of $L$ elements are different and the more the element is in the middle, the larger the proportion coefficient is. Correspondingly, the more the element is at the two ends, the smaller the proportion coefficient is. As is shown in Fig \ref{fig6}(a), the big value of bucket 13 is close to the high passband of Dirichlet kernel filter ($(1660+64)/128-13\approx0.47\approx 0.5$), so its proportion coefficient is large, then the location and estimation of this bucket is very easy. Correspondingly, the proportion coefficients of other big values in other buckets are small. As is shown in Fig \ref{fig6}(b), the proportion coefficient of the big value in bucket 0 is large ($(64+0)/128-0=0.5\approx 0.5$), and others are small. As is shown in Fig \ref{fig10}(a), for the big value $\hat x _{610}$ in bucket 5, although the amplitude of significant frequency coefficients in some filtered signals is reduced due to the relationship of the proportional coefficient, as long as it is still larger than the energy of noise multiplied by larger scale coefficient, the position of them can also be correctly located.

\section*{The second stage of sFFT: Spectrum reconstruction}
Through the first stage of sFFT, bucketization, $N$ frequencies are divided into $B$ buckets by one of the three filters introduced in the previous chapter. The next stage of sFFT is the spectrum reconstruction by identifying frequencies that are isolated in their buckets. The operation is generally divided into two steps. The first step is the location, and the second step is the estimation. If the location and estimation of the frequency of all the large value scattered into the bucket can be obtained correctly, the purpose of sFFT is achieved through the sum of these results. It is clear that buckets can be classified into three categories: 'zero-ton' bucket, 'single-ton' bucket, 'multi-ton' bucket. There is a big difference in the way to solve the 'single-ton' bucket or the 'multi-ton' bucket. 
\subsection*{The method of locating the position}
\subsubsection*{Phase encoding method}
The shift of time leads to the phase shift of frequency. Suppose the bucket $i$ is a 'single-ton' bucket we can obtain $\hat{y}_i \approx \text{prony} (\hat{x}_{f_i})\omega ^ {0 \cdot {f_i}})$ with the time shift operation of $\mathbf{S}_{\tau} = \mathbf{S}_{0}$ (while $\hat{x}_{f_i}$ is the only significant frequency coefficient in the bucket $i$) for the first time. Then we can obtain $\hat{y'}_i\approx \text{prony}(\hat{x}_{f_i})\omega ^ {1 \cdot {f_i}}$ with the time shift operation of $\mathbf{S}_{\tau} = \mathbf{S}_{1}$ in the second time. The position $f_i$ can be obtained by $ f_i=\angle \frac{\hat{y'}_i}{\hat{y}_i}\cdot (\frac{-N}{2\pi})$. 

The advantage of this method is that it is easy to use and has low runtime complexity and sampling complexity. The disadvantage is that once the noise in the bucket is large, it is easy to affect the judgment of position, and this method can only deal with the condition that the default is 'single-ton' bucket.

\subsubsection*{The Method of probability and statistics}
The method of probability statistics is to judge whether the frequency position is effective or invalid by counting the probability of each original frequency position appearing in the large value bucket and the small value bucket. The details are as follows: In one statistic, after bucketization, in $B$ buckets, if $\hat {y}_{i_1},\hat {y}_{i_2},\dots,\hat {y}_{i_{2K}}$ are large, it means that positions of the effective spectrum may appear in the frequency elements that are hashed by these buckets. Correspondingly, $\hat {y}_{i_{2K+1}},\hat {y}_{i_{2K+2}},\dots,\hat {y}_{i_B}$ are small; it means that positions of the effective spectrum does not exist in the frequency elements that are hashed by these buckets. After a large number of randomly bucketization and statistical voting, the locations of the high probable probability is the target we want to know.

The advantage of this method is high robustness and can deal with the condition of 'multi-ton' bucket, but the disadvantage is that it is a probabilistic algorithm. If we want to improve the success probability, it will increase runtime complexity and sampling complexity.

\subsubsection*{The Method of binary search or multi-scale search}
This method needs to subdivide the separated bucket iteratively. By reducing the location range, the final position can be located successfully. The application of a binary search is described as follows. If the bucket consists of $L$ neighboring elements, suppose the bucket $i$ is a 'single-ton' bucket and $f_i$ is the only effective position, we can obtain $\hat{y}_i=\text{prony}(\hat{x}_s,\hat{x}_{s+1},\dots,\hat{x}_{s+L-1})$ at first ($s$ represents the starting position of adjacent elements, and $s+L-1$ represents the end position of adjacent elements). Through the second sub-bucketization by subsampled in the frequency domain, we can obtain $\hat{y}_{i_{11}}=\text{prony}(\hat{x}_s,\hat{x}_{s+2},\hat{x}_{s+4}\dots)$ and $\hat{y}_{i_{12}}=\text{prony}(\hat{x}_{s+1},\hat{x}_{s+3},\hat{x}_{s+5}\dots)$. If $\hat{y}_{i_{11}}>\hat{y}_{i_{12}}$, then $r_0=0$($r_0$ represents the result of the first binary search), so $f_i \text{mod} 2 = 0 = r_0$. If $\hat{y}_{i_{11}}<\hat{y}_{i_{12}}$, then $r_0=1$, so $f_i \text{mod} 2 = 1 = r_0$. In the next search, if $r_0=0$, subdivide $\hat{y}_{i_{11}}$, we can obtain $\hat{y}_{i_{21}}=\text{prony}(\hat{x}_s,\hat{x}_{s+4},\hat{x}_{s+8}\dots)$ and $\hat{y}_{i_{22}}=\text{prony}(\hat{x}_{s+2},\hat{x}_{s+6},\hat{x}_{s+10}\dots)$. If $r_0=1$, subdivide $\hat{y}_{i_{12}}$, we can obtain $\hat{y}_{i_{21}}=\text{prony}(\hat{x}_{s+1},\hat{x}_{s+5},\hat{x}_{s+9}\dots)$ and $\hat{y}_{i_{22}}=\text{prony}(\hat{x}_{s+3},\hat{x}_{s+7},\hat{x}_{s+11}\dots)$. If $\hat{y}_{i_{21}}>\hat{y}_{i_{22}}$, then $r_1=0$($r_1$ represents the result of the second binary search). If $\hat{y}_{i_{21}}<\hat{y}_{i_{22}}$, then $r_1=1$, so $f_i \text{mod} 4 = 2r_1 + r_0$. In the third search, if $r_1=0$, $\hat{y}_{i_{31}}$ and $\hat{y}_{i_{32}}$ are obtained by dividing $\hat{y}_{i_{21}}$, On the contrary, they are obtained by dividing $\hat{y}_{i_{22}}$. The subsequent processes are similar. At last we obtain $f_i \text{mod} L = r_0 +2r_1 + 4r_2 + 8r_3 + \dots$. The method of multi-scale search is similar; the difference between them is the number of subdividing is from two to $l$(let $l$ be multi-scale parameters). The process of narrowing down the scope is changed from $L \to L/2 \to L/4 \to \dots$ in the binary search to $L \to L/l \to L/(l^2) \to \dots$ in the multi-scale search.

The advantage of this method is that it is medium robustness and has good runtime complexity. The disadvantage is that this method can only deal with the case of 'single-ton' bucket by default and the runtime complexity and sampling complexity are affected by $l$ when $L$ has been determined.

\subsubsection*{The method of Prony}
The above three methods can solve the case of 'single-ton' bucket by default. The Prony method can deal with 'multi-ton' bucket. In view of the similarity of the bucket structures of the three filters, the filtered signal through the spike train filter is taken as an example.

Firstly calculate $\hat{y}_{B,\tau}=\mathbf{F}_{B}\mathbf{D}_{L}\mathbf{S}_{\tau}x$ representing filtered spectrum through bucketization. Suppose in bucket $i$, the number of significant frequencies is denoted by $a$; we get simplified Equation (\ref{Eq23}) from Equation (\ref{Eq21}) and Equation (\ref{Eq22}). In Equation (\ref{Eq22}) and Equation (\ref{Eq23}), $p_j=\hat{x}_{f_{j}}$ respecting effective frequency values for $\left | p_0 \right |\geq \left | p_1 \right |\geq \dots\geq \left | p_{a-1} \right |\gg \text{other values of } \ | p_j  \ |$, $z_j=\omega^{f_j}$ respecting effective frequency position for $f_j \in \left \{ i,i+L,\dots,i+(L-1)B) \right \}$, $m_k=\hat y _{B,k }[i]$ respecting filtered spectrum in bucket $i$ by different time shift parameters $k=\{0,1,..,2a-1\}$. 

\begin{equation}
\hat y _{B,\tau }[i]=\sum_{j=0}^{L-1}{\hat {x} _{jB+i} \omega ^{\tau(jB+i)}}
\approx \sum_{j=0}^{a-1}{\hat {x}_{f_j}}\omega ^{ \tau f_j} \label{Eq21}
\end{equation}
\begin{equation}
m_k  \approx \sum_{j=0}^{a-1}p_j z_j^{k} \label{Eq22}
\end{equation}
\begin{equation}
\begin{bmatrix}
z_0^{0}& z_1^{0} & \cdots &z_{a-1}^{0} \\ 
z_0^{1}& z_1^{1} & \cdots &z_{a-1}^{1} \\ 
\cdots & \cdots & \cdots & \cdots\\ 
z_0^{2a-1}& z_1^{2a-1} & \cdots &z_{a-1}^{2a-1}
\end{bmatrix}
\begin{bmatrix}
p_0\\ 
p_1\\
\cdots\\
p_{a-1} 
\end{bmatrix}
\approx 
\begin{bmatrix}
m_0\\ 
m_1\\
\cdots\\
m_{2a-1} 
\end{bmatrix} \label{Eq23}
\end{equation}

The first thing is to obtain the number of significant frequencies in bucket $i$. The method is to calculate the largest $a_m$ principal components in each bucket(suppose there are at most $a_m$ number of significant frequencies aliasing in every bucket). Among the total $a_m B$ number of principal components, the first $2B$ principal components may be the target. How many numbers of the $a_m$ principal components in a bucket are the large values of all principal components is the number of collisions in this bucket. The principal component decomposition (PCA) method is to symmetric singular value decomposition (SSVD) of the Hankel matrix which is composed of filtered signal groups with interval time shift. After knowing the number of collisions, the aliasing problem is reformulated as Moment Preserving Problem(MPP). The orthogonal polynomial formula $P(z)$ is defined as Equation (\ref{Eq24}) and $P(z) \approx 0$. The Matrix $\mathbf{M}_{a}\in \mathbb{C}^{a \times a}$ is defined as Equation (\ref{Eq25}). The vector $C$ is defined as $C=[c_0,c_1,\dots,c_{a-1}]^T$. The vector $M_s$ is defined as $M_s=[-m_a,-m_{a+1},\dots,-m_{2a-1}]^T$. The MPP problem formulated by Bose-Chaudhuri-Hocquenghem(BCH) codes is expressed by Equation (\ref{Eq26}). Through the moments’ formula $\mathbf{M}_{a} C \approx M_s$, we can obtain $C \approx (\mathbf{M}_{a}) ^ {-1} M_s$. After gaining $C$, there are many ways to obtain $z_j$'s through Equation (\ref{Eq24}) and $P(z) \approx 0$. After knowing $z_j$'s, we can obtain approximate positions $f_j$'s through $z_j=\omega^{f_j}$.
\begin{equation}
P(z)= z^a+c_{a-1}z^{a-1}+ \dots + c_1z+c_0	 \label{Eq24}
\end{equation}
\begin{equation}
\mathbf{M}_{a}=\begin{bmatrix}
m_0 &m_1  &\cdots  & m_{a -1}\\ 
m_1 &m_2  &\cdots  & m_{a }\\ 
\cdots & \cdots & \cdots & \cdots\\ 
m_{a -1} &m_{a_m}  &\cdots  & m_{2a -2}
\end{bmatrix}_{{a}\times {a}} \label{Eq25}
\end{equation}
\begin{equation}
\begin{bmatrix}
m_0 &m_1  &\cdots  & m_{a -1}\\ 
m_1 &m_2  &\cdots  & m_{a }\\ 
\cdots & \cdots & \cdots & \cdots\\ 
m_{a -1} &m_{a_m}  &\cdots  & m_{2a -2}
\end{bmatrix}_{{a}\times {a}}\begin{bmatrix}
c_0\\ 
c_1\\ 
\dots\\ 
c_{a-1}
\end{bmatrix}_{a \times 1}\approx \begin{bmatrix}
-m_a\\ 
-m_{a+1}\\ 
\dots \\ 
-m_{2a-1}
\end{bmatrix}_{a \times 1} \label{Eq26}
\end{equation}

The advantage of this method is that it can solve all types of buckets by one-shot and has good sampling complexity, but the disadvantage is that there is a premise that at most $a_m$ number of significant frequencies aliasing in every bucket and the method needs matrix operation, which is time-consuming.

\subsection*{The method of estimating the value}
When we get the position by the first step, in the case of spike train filter and flat filter, the subsequent estimation is relatively simple. As to the Dirichlet kernel filter, the estimation can be carried out with higher probability and higher accuracy by using the property of random sampling.
\subsubsection*{Formula method}
If the frequency set of known large value is $f= \{ f_1,f_2, \dots f_k \}$ and $K$ is very small, the expression of Fourier series of $K$ frequencies can be obtained according to Equation (\ref{Eq2}) directly. As long as $K \ll \text{log}N$, its operation efficiency is acceptable.
\subsubsection*{Energy concentration method}
In the case of spike train filter and flat filter, it is relatively simple to estimate the value of the only one large value frequency coefficient of the 'single-ton' bucket which has been successfully located. Because the passband length of these two filters is approximately equal to $L$, the spike train filter has no spectrum leakage; the flat filter basically has no spectrum leakage. Therefore, in order to simplify and facilitate to do the component analysis of the bucket, other components are ignored in one bucket. So in simple words, the energy of the bucket is viewed as only contributed by the large value frequency coefficient of 'single-ton' bucket. For the flat filter, the bucket is composed of adjacent elements, so we can obtain $\hat{y}_{i}=\text{prony}(\hat{x}_s,\hat{x}_{s+1},\hat{x}_{s+2}\dots,\hat{x}_{s+L-1})$ and $\hat{y}_{i} \approx  \text{prony}(\hat{x}_{f})$ ($\hat{x}_{s}$ is the first element in the bucket, $\hat{x}_{f}$ is the located element in the bucket). For the spike train filter, the bucket is composed of $L$ elements with the same remainder divided by $B$, so we can obtain $\hat{y}_{i}=\text{prony}(\hat{x}_s,\hat{x}_{s+B},\hat{x}_{s+2B}\dots,\hat{x}_{s+(L-1)B})$ and $\hat{y}_{i} \approx  \text{prony}(\hat{x}_{f})$. Through $\hat{y}_{i} \approx  \text{prony}(\hat{x}_{f})$, we can obtain $\hat{x}_{f}$ which we wanted; and the average value of multiple buckets is more accurate.
\subsubsection*{Method of original signal frequency shift}
\begin{figure} [!ht]\centering  
\subfigure[The original signal and filtered signal.]{  \includegraphics[width=0.45\columnwidth]{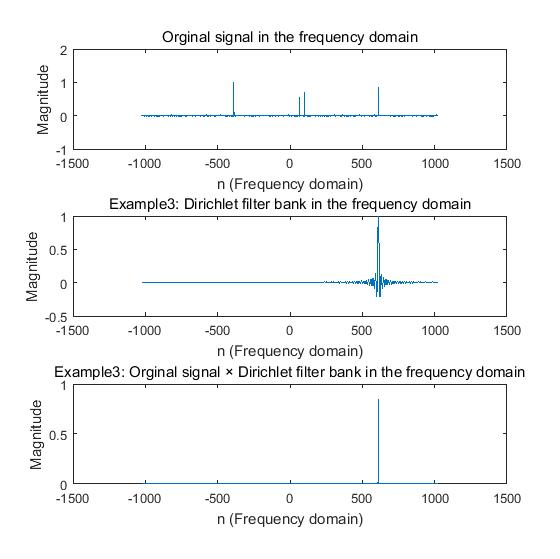} } \subfigure[The new signal and filtered signal.] { \includegraphics[width=0.45\columnwidth]{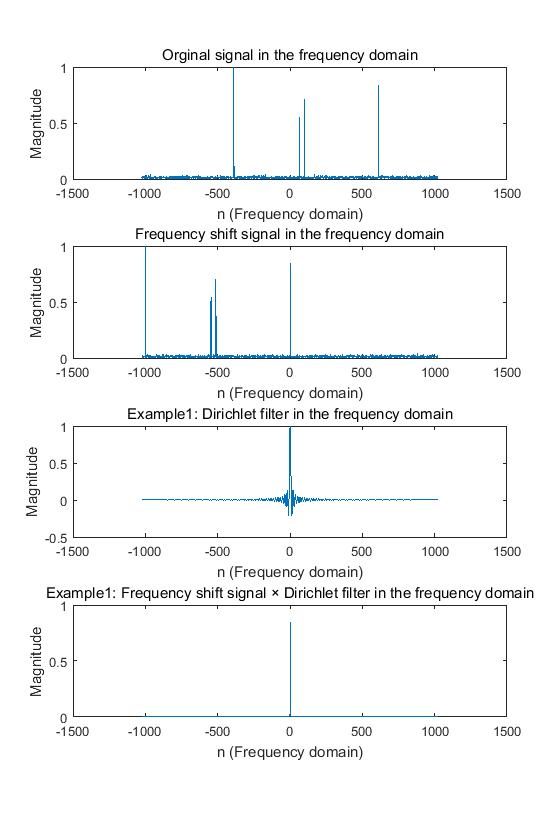} } 
\caption{The original signal, new signal after frequency shift, filtered signal in the frequency domain.}  \label{fig7} 
\end{figure}

In the case of the Dirichlet kernel filter, it is necessary to move the frequency coefficient located to the frequency position is equal to zero. If a vector $x'=\mathbf{V}_{f} x$, such that $x'_i=x_{ i}\omega _{N}^{fi}$ ($f$ is the located position in the bucket) and $\hat{x'}_{0}=\frac{1}{N}(x'_0+x'_1+x'_2+\dots+x'_{N-1})=\hat{x}_{f}$. Let $T$ be the random variable that takes each value from set $\{ 0,1,2, \dots , N-1   \}$ with equal probability. Since the new signal after frequency shift has only one large frequency coefficient $\hat{x'}_{0}$, if sampled randomly, it satisfies $\frac{1}{N}(x'_0+x'_1+x'_2+\dots+x'_{N-1})\approx \frac{1}{t}(x'_{T_0}+x'_{T_1}+x'_{T_2}+\dots+x'_{T_{t-1}})$. Therefore, the random sampling and averaging of the signal $x'_{T_0},x'_{T_1},x'_{T_2},\dots,x'_{T_{t-1}}$ is enough to calculate $\hat{x'}_{0}$ and $\hat{x}_{f}$. The example is shown in Fig \ref{fig7}.

\subsubsection*{The method of Prony}
Suppose in bucket $i$; if the number of significant frequencies $a$ and approximate effective frequency position $z_j$'s have been known, we can get Equation (\ref{Eq27}) for $P$ number of random sampling with time shift parameter $r_1,\dots,r_P$. The solution to obtain $p_j$'s is the subspace pursuit method by random sampling according to the properties of the sparse matrix and the idea of greedy pursuit by using compressed sensing (CS) concept.
\begin{equation}
\begin{bmatrix}
m_1\\ 
\dots\\ 
m_P
\end{bmatrix}_{P\times 1}=
\begin{bmatrix}
z_{0}^{r_{1}} &\dots  & z_{L-1}^{r_1}\\ 
 \dots& \dots & \dots\\ 
z_{0}^{r_{P}} & \dots & z_{L-1}^{r_{P}}
\end{bmatrix}_{P\times L} 
\begin{bmatrix}
p_0\\ 
\dots \\ 
p_{L-1}
\end{bmatrix}_{L\times 1}\approx 
\begin{bmatrix}
z_{0}^{r_{1}} &\dots  & z_{a-1}^{r_1}\\ 
 \dots& \dots & \dots\\ 
z_{0}^{r_{P}} & \dots & z_{a-1}^{r_{P}}
\end{bmatrix}_{P\times a} 
\begin{bmatrix}
p_0\\ 
\dots \\ 
p_{a-1}
\end{bmatrix}_{a\times 1} \label{Eq27}
\end{equation}

\section*{The frameworks and performance of sFFT algorithm}
In the above two chapters, two stages of sFFT are introduced in detail. The content includes specific methods and practical examples of bucketization and spectrum reconstruction. This chapter first introduces the problems that be caused by bucketization, then introduce three methods to solve these problems. Finally, three main frameworks are summarized, and the theoretical performance of their corresponding algorithms is analyzed.
\subsection*{The problems caused by bucketization}
Bucketization means to divide multiple frequency coefficients into one bucket. Obviously, this operation will cause two problems. The first problem is frequency aliasing. Once aliasing occurs, 'multi-ton' bucket will be generated. For 'multi-ton' bucket, some location methods and estimation methods are powerless. The second problem is spectrum leakage, which is determined by the nature of the filter, which will lead to positioning error and greatly reduce the accuracy of estimation, even in the case of 'single-ton' bucket.
\subsubsection*{Frequency aliasing}
\begin{figure} [!ht]\centering  
\subfigure[An example of frequency aliasing.]{  \includegraphics[width=0.45\columnwidth]{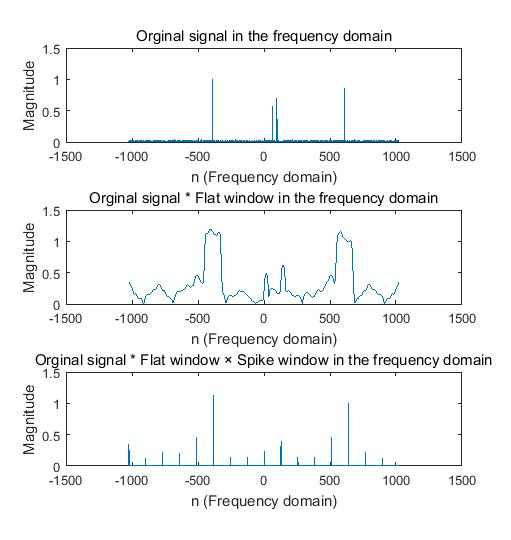} } \subfigure[An example of frequency aliasing.] { \includegraphics[width=0.45\columnwidth]{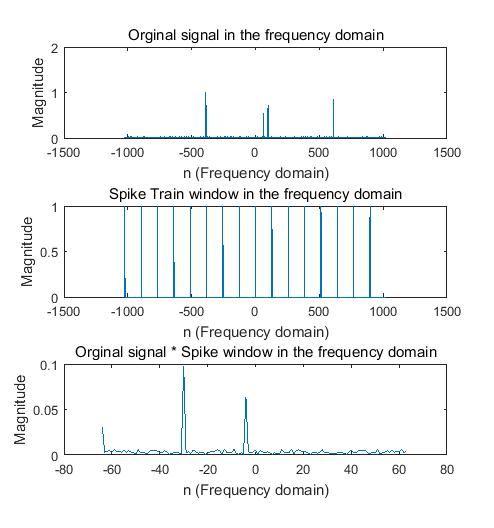} } 
\caption{Two examples of frequency aliasing.}  \label{fig8} 
\end{figure}

As can be seen from the previous chapter, in order to realize hash mapping and reflection mapping, the bucket is composed of either adjacent frequency coefficients or separated frequency coefficients. Once there are two or more large values in one bucket, a 'multi-ton' bucket will be generated. As is shown in Fig \ref{fig8}, the original signal of this experiments is considered as a $N(N = 2048)$ size signal that has only $K(K = 4)$ significant coefficients $\hat x _{64}=0.55, \hat x _{98}=0.7, \hat x _{610}=0.85, \hat x _{1660}=1.0$ while the rest of the coefficients are white Gaussian noise approximately equal to zero. Since position 64 and position 98 are adjacent, the 'multi-ton' bucket will be generated when the signal is filtered by the flat filter in the frequency domain. Since position 98 and 610 has the same remainder over the number 128 (98 mod 128 = 98, 610 mod 128 =98), the 'multi-ton' bucket will be generated when the signal is filtered by the spike train filter in the frequency domain.

\subsubsection*{Spectrum leakage}
Whether the spectrum is leaked or not is based on the nature of the filter. In the energy of a bucket, the proportion coefficients of $L$ elements are different, and the more the element is in the middle, the larger the proportion coefficient is. Correspondingly, the more the element is at both ends, the smaller the proportion coefficient is. If the coefficients on both ends are too small or almost no, it will cause spectrum leakage. As is shown in Fig \ref{fig1}, Fig \ref{fig3}, Fig \ref{fig5}, filtered signal through spike train filter has no spectrum leakage, filtered signal through flat filter has almost no spectrum leakage except both ends of the passband, filtered signal through Dirichlet kernel filter bank has large spectrum leakage except the center of the passband.

\subsection*{The methods to solve these problems}
\subsubsection*{Scaling method (not applicable to spike train filter)}
\begin{figure} [!ht]\centering  
{  \includegraphics[width=0.95\columnwidth]{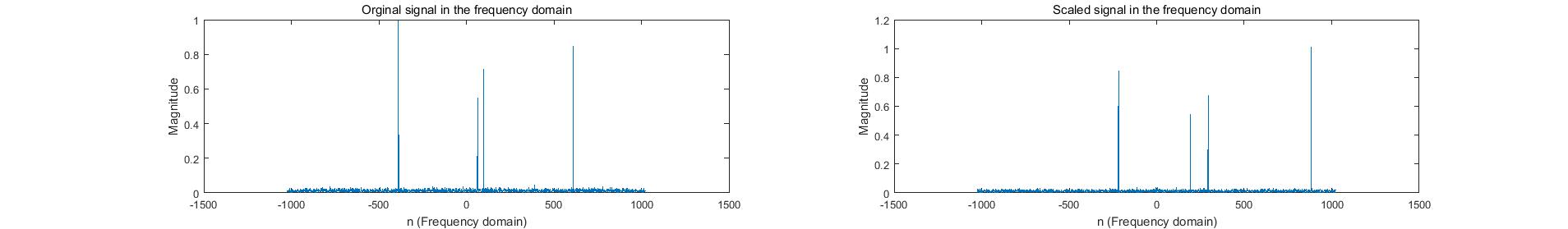} } 
\caption{The original signal and the scaled signal in the frequency domain.}  \label{fig9} 
\end{figure}

The scaling change in the time domain will lead to the corresponding scaling change in the frequency domain. It can make the adjacent frequency no longer close to each other, and the result of the scaled signal obtained will return to the original frequency easily. The scaling change does not change the remainder result, so scaling change is not useful for the spike train filter. An example is shown in Fig \ref{fig9}; it can be seen that the original set of large frequency coefficients is changed from set $\{ 64,98,610,1660 \}$ to a new set $\{ 192(192=64*3), 294(294=98*3), 1830(1830=610*3), 884(884=1660*3 \text{mod} 2048) \}$ through scaling change. The position 64 and position 98 become position 192 and position 294 which are no longer close together, so the frequency coefficients in the bucket are no longer aliased.

\subsubsection*{The method of bucketizations with different parameters}
The second method to deal with 'multi-ton' bucket is to convert 'multi-ton' bucket to 'single-ton' bucket by by using a layer-by-layer peeling method(subtracting the obtained elements). We require the size $N$ of signal $x$ to be a product of a few(typically three or more) co-prime numbers $p_i$'s. For example $N=p_{1} p_{2} p_{3}$ where $p_1, p_2, p_3$ are co-prime numbers. We can use different parameters(e.g. $B_1=p_1, B_2=p_2, B_3=p_3$) to make bucketizations. In the case of these cycles, when a bucket is a 'single-ton' in one cycle, the position and value of this one significant frequency coefficient in this bucket can be easily obtained. This result can be used to peel off the 'multi-ton' bucket in another cycle so that the known information can help it may change from 'multi-ton' bucket to 'single-ton' bucket in the next layer. For example, Consider a $N(N = 20)$ size signal that has only $K(K = 5)$ significant coefficients $\hat x _1, \hat x _3, \hat x _5, \hat x _{10}, \hat x _{13} \gg 0$, while the rest of the coefficients are approximately equal to zero. With this precondition, there are two bucketization cycles. In the first cycle, for $B_1=4$ and $L_1=5$, we get four vectors $ \{ \overrightarrow{y}_{4,0}, \overrightarrow{y}_{4,1}, \overrightarrow{y}_{4,2}, \overrightarrow{y}_{4,3} \} $ respecting the filtered spectrum in four buckets. In the second cycle, for $B_2=5$ and $L_2=4$, we get five vectors $ \{ \overrightarrow{y}_{5,0}, \overrightarrow{y}_{5,1}, \overrightarrow{y}_{5,2}, \overrightarrow{y}_{5,3} , \overrightarrow{y}_{5,4} \} $. $\overrightarrow{y}_{4,2}$ is a 'single-ton' bucket which only has one big value $\hat x _{10}$, and $\overrightarrow{y}_{5,0}$ is a 'multi-ton' bucket which has two big value $\hat x _{10}$ and $\hat x _{5}$. The $\overrightarrow{y}_{5,0}$ can be changed to a 'single-ton' bucket by subtract $\hat x _{10}$ which is obtained by $\overrightarrow{y}_{4,2}$.

The advantage of this method is that it can solve all buckets through peeling layer by layer and has good sampling complexity, but the disadvantage is that this method needs to add a step to determine the type of bucket. In the extreme case of 'multi-ton' bucket, it takes extra time to peel in several layers. By use of graph theory, the efficiency of peeling can also be improved.

\subsubsection*{The method of filter frequency shift operation}
\begin{figure} [!ht]\centering  
\subfigure[Part of the filtered signal.]{  \includegraphics[width=0.45\columnwidth]{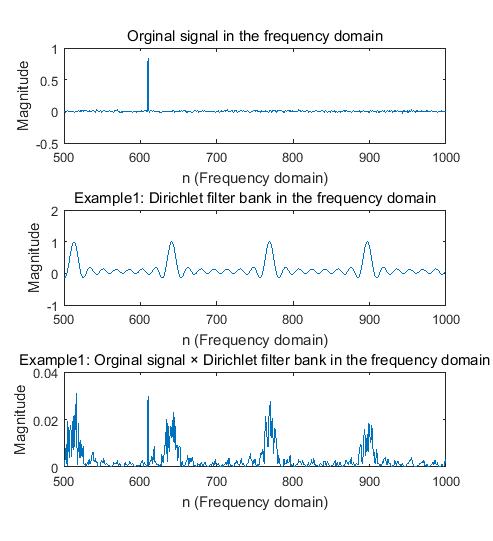} } \subfigure[Part of the filtered signal changed.] { \includegraphics[width=0.45\columnwidth]{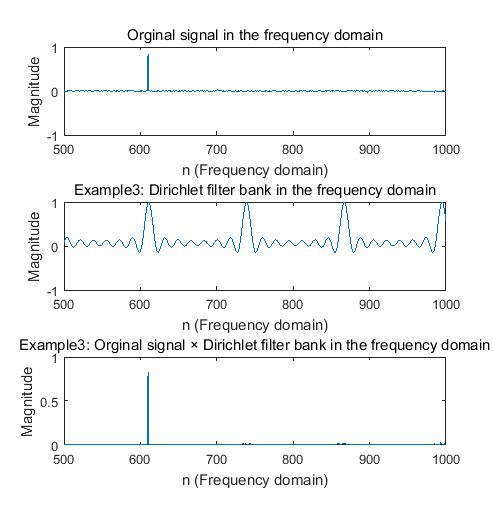} } 
\caption{The original signal and filtered signal in the frequency domain after filter shift.}  \label{fig10} 
\end{figure}

Filter Frequency shift operation can change the center position of the passband, which can make the filtered signal no longer have frequency leakage, even after bucketization through Dirichlet kernel filter bank. As shown in Fig \ref{fig10}(a), the central passband of the original filter is not close to the large value coefficient $\hat{x}_{610}$ of the original frequency. It leads to spectrum leakage when analyzing the filtered signal. However, as shown in Fig \ref{fig10}(b), the position of the center passband of the new filter is changed by frequency shift with the calculated value. In this case, the central passband of the filter is close to the large value coefficient of the original frequency, and the problem of spectrum leakage in the filtered signal can be overcome by this method.

\subsection*{Four algorithmic frameworks}
In the above contents, this paper introduces two stages of sFFT and the technology involved, then summarizes the treatment methods for the frequency aliasing and spectrum leakage problems caused by bucketizations. Next, we will introduce the algorithmic framework for comprehensively utilizing these technologies and methods.

\subsubsection*{The framework of one-shot}
\begin{figure} [!ht]\centering  
{  \includegraphics[width=0.95\columnwidth]{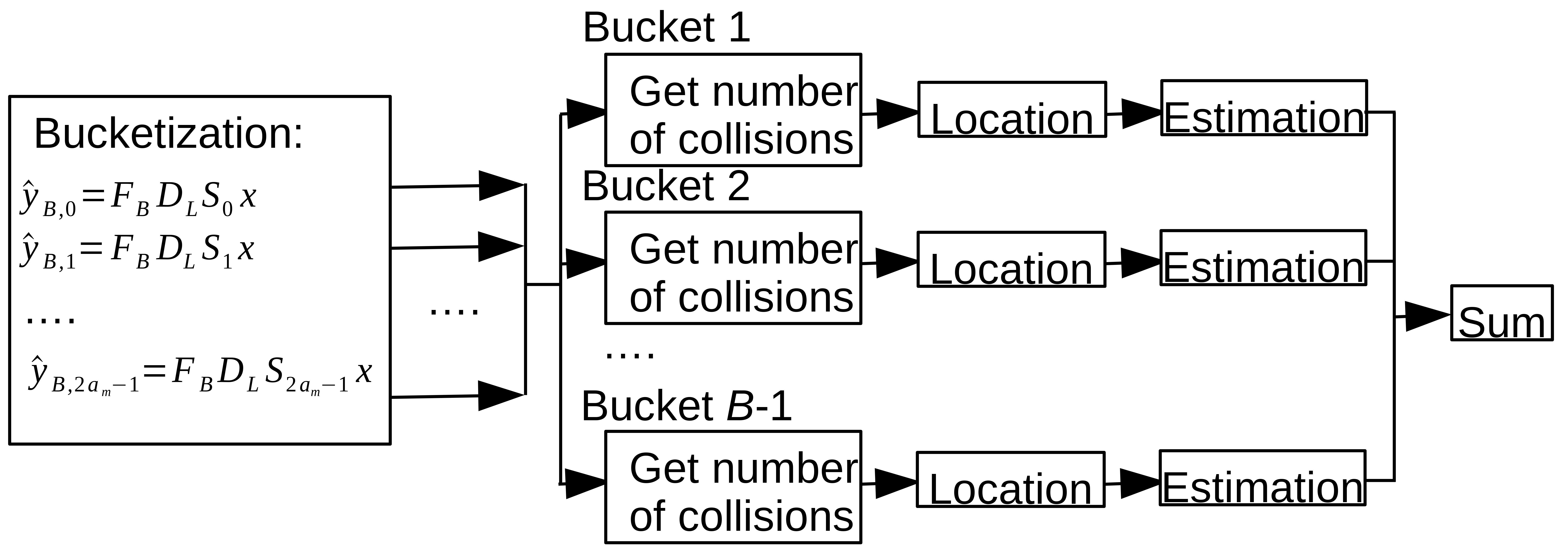} } 
\caption{The system block diagram of the one-shot framework.}  \label{fig11} 
\end{figure}

The idea of the one-shot framework is a sequential process to the end. The block diagram of the one-shot framework is shown in Fig \ref{fig11} and the algorithms based on this framework are shown in Table \ref{tab1}. The advantages of this framework are as follows: 1) It does not need to bucketizations too many times, so it has low sampling complexity. 2) There is no need for random bucketization and iterative process; the extra computation can be avoided. 3) Because of the use of the spike train filter, there is no spectrum leakage. The disadvantages of this framework are as follows: 1) The algorithm framework has a presupposition that the maximum number of aliasing in each bucket is not greater than $a_m$, so the algorithm based on this framework is not a deterministic algorithm. 2) When $a_m$ is set to a large value, the SSVD decomposition or Matrix operation used in this framework is computationally expensive and time-consuming.
\subsubsection*{The framework of voting}
\begin{figure} [!ht]\centering  
{  \includegraphics[width=0.95\columnwidth]{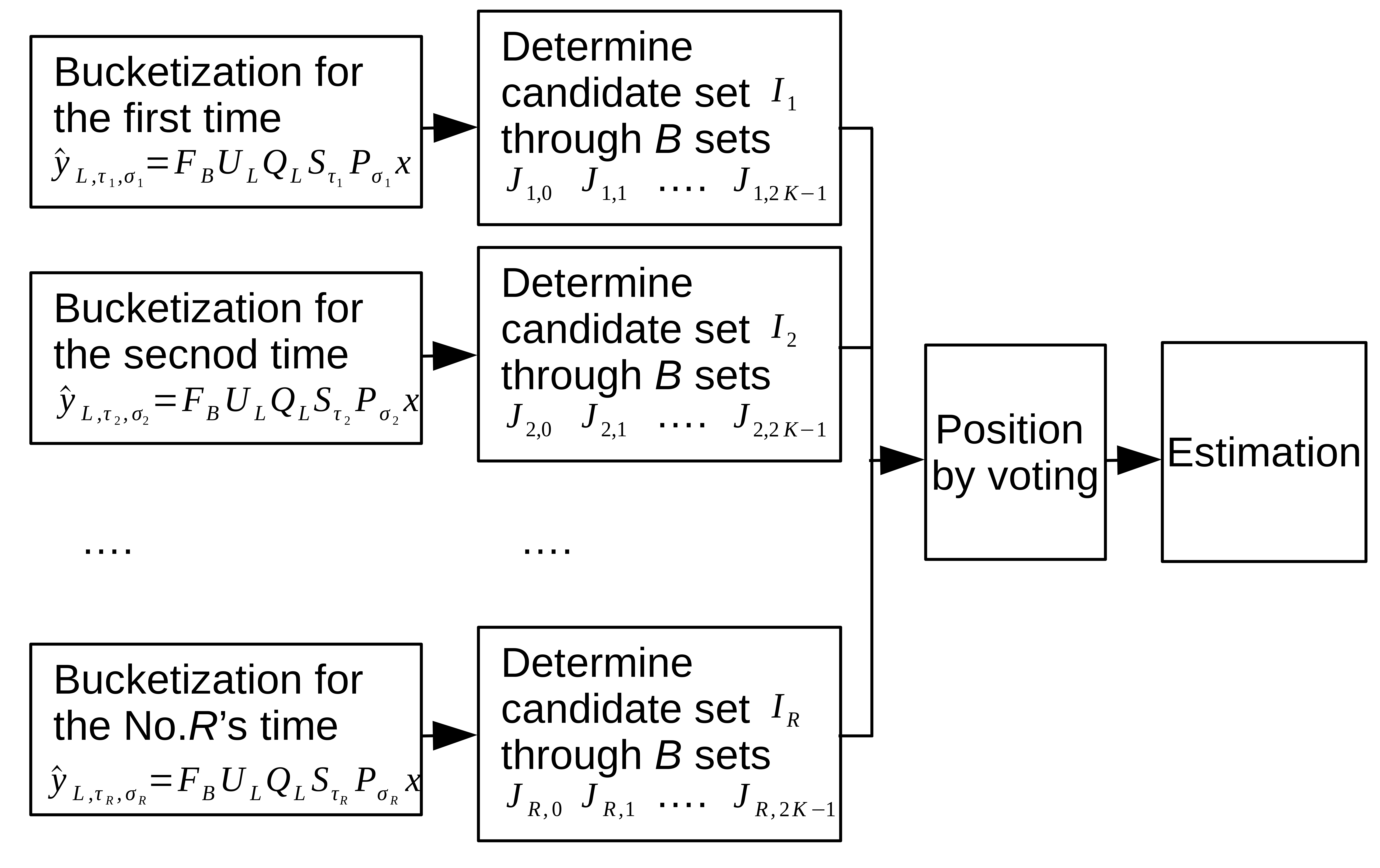} } 
\caption{The system block diagram of the voting framework.}  \label{fig12} 
\end{figure}

The bucketizations of the voting framework are bucketizations after random frequency scaling operation; it can scatter the frequency with a high probability in each bucketizations. It uses the idea of statistics in the steps of positioning. The block diagram of the voting framework is shown in Fig \ref{fig12}, and the algorithms based on this framework are shown in Table \ref{tab1}. The advantages of this framework are as follows: 1) The framework is easy to implement because it does not need iteration, it only needs to count and vote to determine the target location after bucketizations. 2) Because the length of support set based on flat filter is short($\ll N$) in the time domain, the sampling and runtime complexity of the framework are acceptable. The disadvantages of this framework are as follows: 1) In extreme cases, a  small value frequency coefficient may be recognized as a big value frequency coefficient in each voting statistics, resulting in misjudgment, so the algorithm based on this framework is not a deterministic algorithm. 2) The number of samples and the amount of calculation needed for each bucketizations is relatively large, which is determined by the support set of the flat filter in the time domain ($ \gg  O(L)$), so its performance is not optimal.
\subsubsection*{The iterative framework}
\begin{figure} [!ht]\centering  
{  \includegraphics[width=0.9\columnwidth]{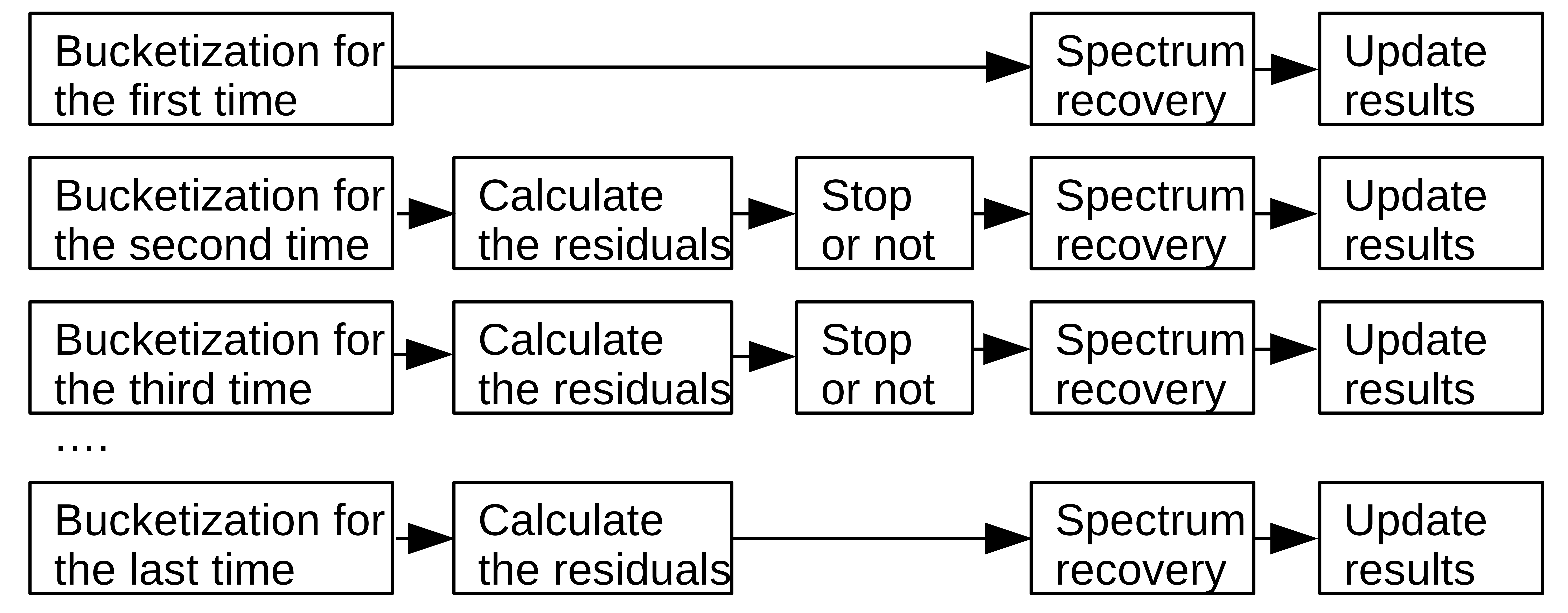} } 
\caption{The system block diagram of the iterative framework.}  \label{fig13} 
\end{figure}

The key point of the iterative framework is that in each new iteration, the contribution of the successfully recovered frequency has been subtracted before spectrum recovery, which helps to deal with some ‘multi-ton’ buckets in spectrum recovery. The block diagram of the iterative framework is shown in Fig \ref{fig13}, and the algorithms based on it are shown in Table \ref{tab1}. The advantages of this framework are as follows: 1) The algorithm based on this framework is a deterministic algorithm. 2) It allows user to take some non-robust but efficient localization methods (such as phase encoding or binary search), because it allows not to complete all spectrum recovery in one-shot, and can approach the optimal result by greedy pursuit in the process of iteration. The disadvantages of this framework are as follows: 1) The framework is more complicated because it takes one more step to subtract contribution. 2) In the worst case, the number of iterations is very large, which affects the efficiency of the algorithm.
\subsubsection*{The peeling framework}
	The key point of the peeling framework is that after bucketizations with different parameters $B$,  the unsolved bucket can be peeled and converted into a solvable bucket by subtracting the evaluated frequency coefficient in the solved bucket. The block diagram of the peeling framework is shown in Fig \ref{fig13}, and the algorithms based on his framework are shown in Table \ref{tab1}. The advantages of this framework are as follows: 1) The algorithm based on this framework is a deterministic algorithm according to the CRT theorem. 2) The complexity of sampling is very low. 3) It has no spectrum leakage. The advantages of this framework are as follows: 1) $N$ has a premise that it is composed of co-coprime numbers, 2) An additional step is needed to determine the character of the bucket. 3) In the worst case, the decoding has to go through several layers of peeling, so the computational complexity becomes high.
\begin{figure} [!ht]\centering  
{  \includegraphics[width=0.9\columnwidth]{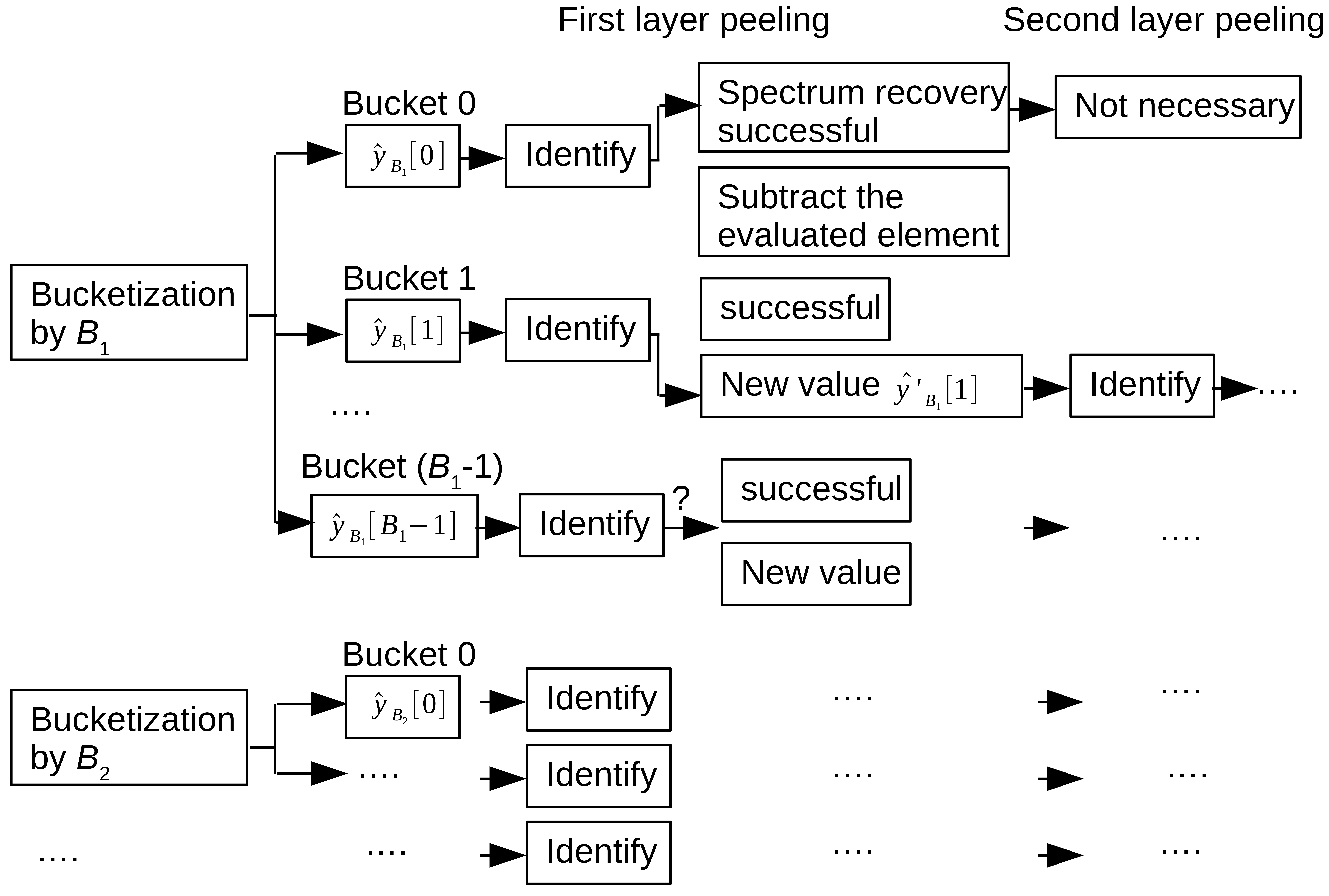} } 
\caption{The system block diagram of the peeling framework.}  \label{fig14} 
\end{figure}	

\subsection*{Theoretical performance of the corresponding algorithm}
Table \ref{tab2} can be concluded with the information of all typical sFFT algorithms and fftw algorithm in theory. Combined with the previous analysis and the contents involved in Table \ref{tab1} and Table \ref{tab2}, we can see the analysis of different types of algorithms are as follows 1) In the comparison of algorithms using different frameworks, the peeling and one-shot framework have better sampling complexity because their sampling complexity is based on $O(B)$. The voting and iterative framework have better runtime complexity because they can use some simple methods in the step of location. 2) As to the positioning methods, the phase encoding method is the most efficient but has poor robustness; the Prony method is the most complicated but has good robustness. 3) As to the estimation methods, the energy concentration method is the most efficient but has poor accuracy; the Prony method is the most complicated but has good accuracy. 3) As to the filters, the algorithms using the spike train filter have good sampling because of their very short support set. The algorithms using the Dirichlet kernel filter bank are the most complicated but can random sampling on account of their estimation method. The algorithms using the flat filter have good performance based on low spectrum leakage and they are easy to use with short support set.	
\begin{table}[!ht] 
\caption{The performance of fftw algorithm and sFFT algorithms in theory}
\setlength{\tabcolsep}{4pt}
\begin{tabular}{|p{60pt}|p{125pt}|p{130pt}|p{45pt}|}
\hline
algorithm& 
runtime complexity& 
sampling complexity & robustness \\
\hline
sFFT1.0&
$O(K^{\frac{1}{2}}N^{\frac{1}{2}}\ \text{log} ^{\frac{3}{2}}N)$& 
$N\left(1-\left(\frac{N-w}{N}\right)^{\ \text{log} N}\right)$& 
medium \\

sFFT2.0&
$O(K^{\frac{2}{3}}N^{\frac{1}{3}} \ \text{log} ^{\frac{4}{3}} N)$& 
$N\left(1-\left(\frac{N-w}{N}\right)^{\ \text{log} N}\right)$& 
medium  \\

sFFT3.0&
$O(K \ \text{log} N) $& 
$O(K \ \text{log} N) $& 
none  \\

sFFT4.0&
$O(K \ \text{log} N \ \text{log}_{(l/(q+1))}(N/K) )$& 
$O(K \ \text{log} N \ \text{log}_{(l/(q+1))}(N/K) )$& 
bad  \\

MPFFT&
$O(K \log N \ \text{log} _2 (N/K) )$& 
$O(K \log N \ \text{log} _2 (N/K) )$& 
good  \\

sFFT-DT1.0&
$O(K \ \text{log} K )$& 
$O(K)$& 
none  \\

sFFT-DT2.0&
$O(K \ \text{log} K + N)$& 
$O(K )$& 
medium  \\

FFAST&
$O( K \ \text{log} K) $& 
$O( K ) $& 
none  \\

R-FFAST&
$ O (K \ \text{log} ^{7/3} N)$& 
$O( K \ \text{log} ^{4/3} K) $& 
good  \\

AAFFT&
$O( K \text{poly} ( \ \text{log} N))$& 
$O( K \text{poly} ( \ \text{log} N))$& 
medium   \\

fftw&
$ O( N \ \text{log} N)$& 
$O(N)$& 
good  \\

\hline

\end{tabular}
\label{tab2}
\end{table}

\section*{Experimental Evaluation}
In this section, we evaluate the performance of sFFT algorithms in the general sparse case. The algorithms include AAFFT, sFFT-DT2.0, R-FFAST, sFFT1.0, sFFT2.0, sFFT4.0, MPFFT, fftw algorithm. The codes of the sFFT1.0, sFFT2.0\footnote{The code is available at http://groups.csail.mit.edu/netmit/sFFT/.}, sFFT3.0, MPFFT\footnote{The code is available at https://github.com/urrfinjuss/mpfft.}, sFFT-DT2.0\footnote{The code is available at https://www.iis.sinica.edu.tw/pages/lcs.}, R-FFAST\footnote{The code is available at https://github.com/UCBASiCS/FFAST.}, AAFFT\footnote{The code is available at https://sourceforge.net/projects/aafftannarborfa/.}, fftw\footnote{The code is available at http://www.fftw.org/.} algorithm are already open sources. A series of experiments consists of three parts. The first part is to test the runtime complexity of different algorithms by recording the run time. The second part is to test the sampling complexity of different algorithms by recording the sampling ratio. The third part is to test the robustness of different algorithms by recording $L_1$ error of the calculation result with different SNR signals. This paper only analyzes the general sparse case, which is more representative. As for the exactly sparse case, we can see the website\footnote{https://github.com/zkjiang/-/tree/master/docs/sfft project/experiment data}. All experiments are run on a Linux CentOS computer with 4 Intel(R) Core(TM) i5 CPU and 8 GB of RAM. 

\subsection*{Experimental Setup}
In the experiment, the test signals are gained in a way that $K$ frequencies are randomly selected from $N$ frequencies and assigned a magnitude of 1.0 and a uniformly random phase. The rest frequencies are set to zero in the exactly case or combined with additive white Gaussian noise in the general case, whose variance varies depending on the SNR required. The general sparse case means SNR=20db. The parameters of these algorithms are chosen so that they can make a balance between time efficiency and robustness. In each experiment, the platform can generate a signal with SNR, $K$, $N$ as required. The prepared signal and the value of $N$ and $K$ will be transmitted to different algorithm libraries through a standard interface. The results of the algorithm library, the time and sampling ratio used in operation will also be returned to the platform. These algorithm libraries are generated by modifying the open source code of the algorithms mentioned above. Each test record contains the run time, the sampling proportion, and the $L_0, L_1, L_2$ error between the calculation result and the best result through the algorithm library. The detail of code, data, report are all open source(The R-FFAST algorithm does not pass through this platform because its signal length is required to be the product of several prime numbers). The new testing platform is developed from the old platform\footnote{https://github.com/ludwigschmidt/sft-experiments}. The detail of codes, data, reports are all open sources\footnote{https://github.com/zkjiang/-/tree/master/docs/sfft project}.

\subsection*{The experiments of time complexity performance in the general sparse case}
\begin{figure} [!ht]\centering  
\subfigure[Runtime vs signal size.]{  \includegraphics[width=0.45\columnwidth]{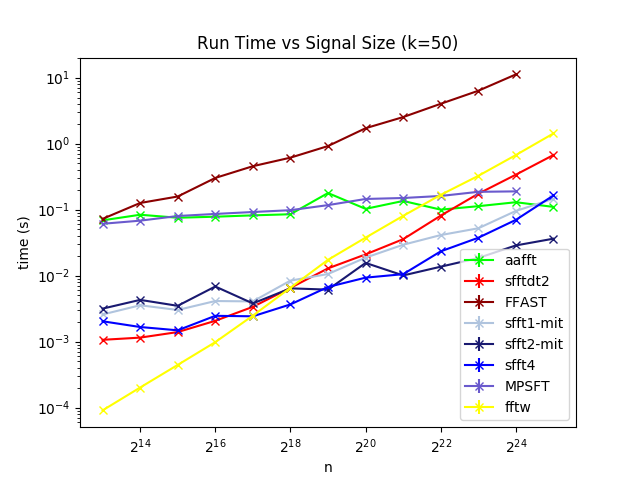} } \subfigure[Runtime vs signal sparsity.] { \includegraphics[width=0.45\columnwidth]{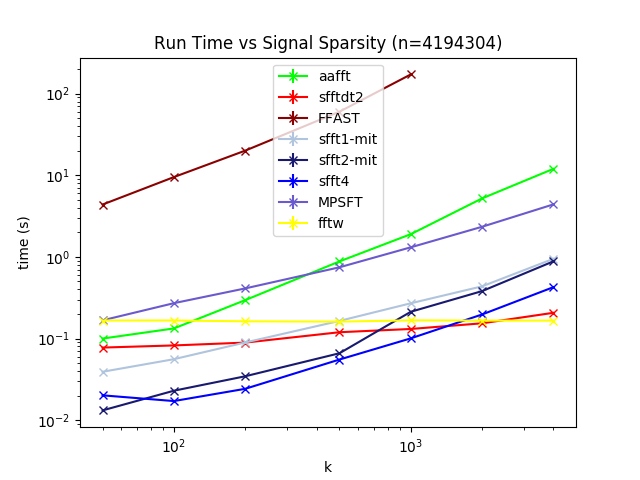} } 
\caption{Runtime of the sFFT algorithm in the general sparse case.}    \label{fig15} 
\end{figure}

We plot Fig \ref{fig15} representing run times vs Signal Size and vs Signal Sparsity for AAFFT, sFFT-DT2.0, R-FFAST, sFFT1.0, sFFT2.0, sFFT4.0, MPFFT, fftw algorithm in the general sparse case. From Fig \ref{fig15}, we can see 1) The run time of these sFFT algorithms are approximately linear in the log scale as a function of $N$ and in the standard scale as a function of $K$. 2) Results of ranking the runtime of eight algorithms is sFFT2.0 $>$ sFFT4.0 $>$ sFFT1.0 $>$ AAFFT $>$ MPSFT $>$ sFFT-DT2.0 $>$ fftw $>$ R-FFAST when $N$ is large. 3) Results of ranking the runtime of eight algorithms is fftw $>$ sFFT-DT2.0 $>$ sFFT4.0 $>$ sFFT2.0 $>$ sFFT1.0 $>$ MPSFT $>$ AAFFT $>$ R-FFAST when $K$ is large. From the ranking, it can be seen that algorithms using the flat filter are the best, algorithms using the Dirichlet kernel filter bank are good, and algorithms using the spike train filter are ordinary in the case of $N$ is large. It also can be seen that algorithms using the spike train are the best, algorithms using the flat filter are good, and algorithms using the Dirichlet kernel filter bank are ordinary in the case of $K$ is large.
	
\subsection*{The experiments of sampling performance in the general sparse case }
\begin{figure} [!ht]\centering  
\subfigure[Signal sampled vs signal size.]{  \includegraphics[width=0.45\columnwidth]{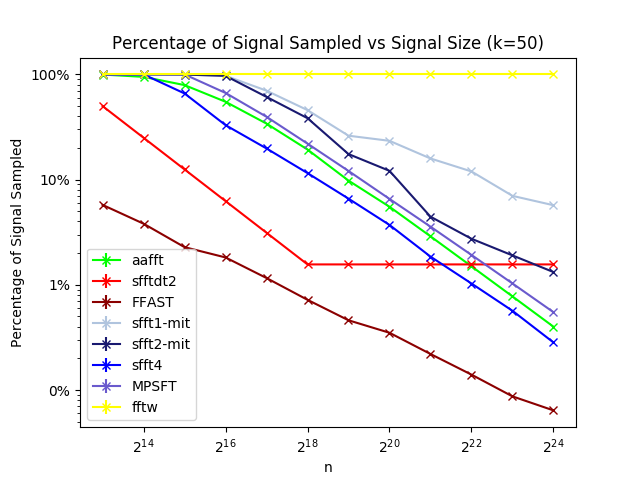} } \subfigure[signal sampled vs signal sparsity.] { \includegraphics[width=0.45\columnwidth]{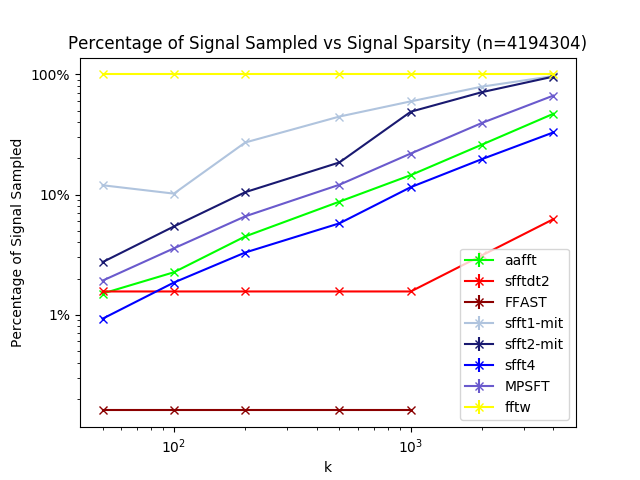} } 
\caption{Percentage of the signal sampled of the sFFT algorithm in the general sparse case.}\label{fig16} 
\end{figure}

We plot Fig \ref{fig16} representing the percentage of the signal sampled vs signal size and vs signal sparsity for AAFFT, sFFT-DT2.0, R-FFAST, sFFT1.0, sFFT2.0, sFFT4.0, MPFFT, fftw algorithm in the general sparse case. From Fig \ref{fig16}, we can see 1) The percentage of the signal sampled of these SFFT algorithms are approximately linear in the log scale as a function of $N$ and in the standard scale as a function of $K$. 2) Results of ranking the sampling complexity of eight algorithms is R-FFAST $>$ sFFT-DT2.0 $>$ sFFT4.0 $>$ AAFFT $>$ MPSFT $>$ sFFT2.0 $>$ sFFT1.0 $>$ fftw. From the ranking, it can be seen that algorithms using the spike train filter are the best, algorithms using the Dirichlet kernel filter bank are good, and algorithms using the flat filter are ordinary.

\subsection*{The experiments of robust performance}
\begin{figure} [!ht]\centering  
\subfigure[Runtime vs SNR.]{  \includegraphics[width=0.45\columnwidth]{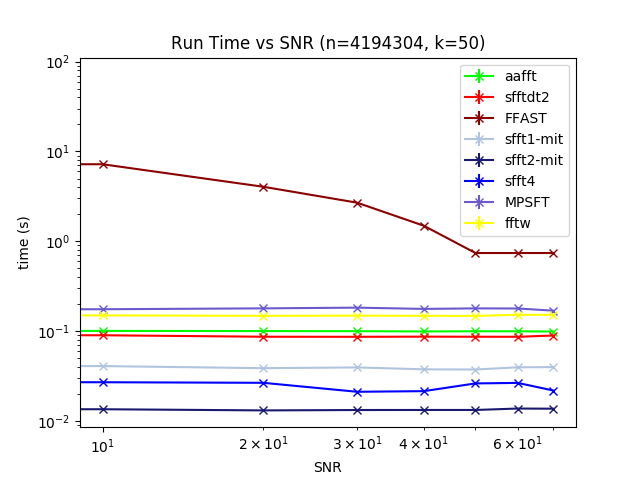} } \subfigure[$L_1$-error  vs SNR.] { \includegraphics[width=0.45\columnwidth]{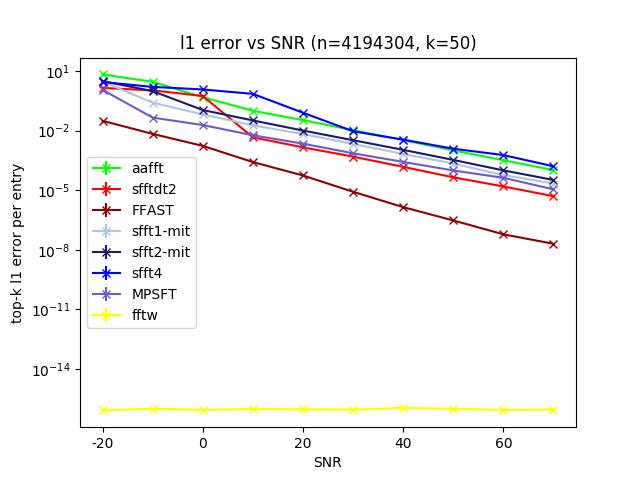} } 
\caption{Runtime and $L_1$-error of the sFFT algorithm in the general sparse case vs SNR}  \label{fig17} 
\end{figure}

We plot Fig \ref{fig17} representing the runtime and $L_1$-error vs SNR for AAFFT, sFFT-DT2.0, R-FFAST, sFFT1.0, sFFT2.0, sFFT4.0, MPFFT, fftw algorithm. From Fig \ref{fig17}, we can see 1) The runtime is approximately equal vs SNR. 2) To a certain extent, these eight algorithms are all robust. 3) Results of ranking the robustness of eight algorithms is fftw $>$ R-FFAST $>$ MPSFT $>$ sFFT1.0 $>$ sFFT2.0 $>$ AAFFT $>$ sFFT4.0. From the ranking, it can be seen that algorithms using the spike train filter are the best, algorithms using the flat filter are good, and algorithms using the Dirichlet kernel filter bank are ordinary.

\section*{Conclusion}
In the first part, the paper gives a detailed introduction to the technology used in sFFT algorithms. When discussing the operation of signal, time shift operation, time scaling operation, frequency shift operation, subsampling operation, aliasing operation are mentioned. When discussing methods of frequency bucketization, flat filter, spike train filter, Dirichlet kernel filter bank are mentioned. As to the methods of location, phase encoding, probability and statistics, Prony, binary search, multiscale search are mentioned. When it comes to methods of estimation, formula, energy concentration, frequency shift, Prony are mentioned. In the second part, the paper gives a detailed introduction to the confusions and frameworks of sFFT. Firstly, it introduces two problems that caused by bucketization, then introduces three methods to solve these problems. Then, four main frameworks are summarized, and the theoretical performance of their corresponding algorithms is analyzed. In the third part, we make three series of experiments to evaluate the performance of all sFFT algorithm in the general case. The experiment includes three parts: runtime complexity, sampling complexity, robustness. The analysis of the experiments satisfies theoretical inference.

The main contribution of this paper is 1) The technical methods used in the stages of sFFT are described in detail, and the corresponding algorithm framework and theoretical performance are also fully analyzed. 2) Develop a standard testing platform that can test more than ten typical sFFT algorithms under all kinds of signal on the basis of the old platform. 3) Get a conclusion of the character and performance of the all sFFT algorithm in theory and practice. Through the summary and analysis of the experimental results, the performance and differences of various technologies used in different algorithms are further verified. These results provide the basis and reference for us to use the characteristics of different algorithms for specific applications.

\section*{Acknowledgments}
This work was supported by Youth Program of National Natural Science Foundation of China under Grant 61703263.

\section*{Author Contributions}

Conceptualization: Zhikang Jiang.

Formal analysis: Bin Li.

Investigation: Bin Li.

Methodology: Zhikang Jiang.

Resources: Zhikang Jiang.

Software: Zhikang Jiang.

Supervision: Bin Li.

Visualization: Zhikang Jiang.

Writing – original draft: Zhikang Jiang, Bin Li.

Writing – review and editing: Zhikang Jiang, Jie Chen.

\nolinenumbers


\begin{thebibliography}{10}

\bibitem{bib1}
Gilbert AC, Guha S, Indyk P, Muthukrishnan S, Strauss M.
\newblock {Near-optimal sparse Fourier representations via sampling}.
\newblock Conference Proceedings of the Annual ACM Symposium on Theory of
  Computing. 2002;2:152--161.
\newblock doi:{10.1145/509931.509933}.

\bibitem{bib2}
Iwen MA, Gilbert A, Strauss M.
\newblock {Empirical evaluation of a sub-linear time sparse DFT algorithm}.
\newblock Communications in Mathematical Sciences. 2007;5(4):981--998.
\newblock doi:{10.4310/cms.2007.v5.n4.a13}.

\bibitem{bib3}
Gilbert AC, Strauss MJ, Tropp JA.
\newblock {A tutorial on fast fourier sampling: How to apply it to problems}.
\newblock IEEE Signal Processing Magazine. 2008;25(2):57--66.
\newblock doi:{10.1109/MSP.2007.915000}.

\bibitem{bib4}
Hsieh SH, Lu CS, Pei SC.
\newblock {Sparse Fast Fourier Transform by downsampling}.
\newblock ICASSP, IEEE International Conference on Acoustics, Speech and Signal
  Processing - Proceedings. 2013; p. 5637--5641.
\newblock doi:{10.1109/ICASSP.2013.6638743}.

\bibitem{bib5}
Hsieh Sh, Lu Cs, Pei Sc.
\newblock {Sparse Fast Fourier Transform for Exactly and Generally}. 2015;.

\bibitem{bib6}
Pawar S, Ramchandran K.
\newblock {Computing a k-sparse n-length Discrete Fourier Transform using at
  most 4k samples and O(k log k) complexity}.
\newblock IEEE International Symposium on Information Theory - Proceedings.
  2013; p. 464--468.
\newblock doi:{10.1109/ISIT.2013.6620269}.

\bibitem{bib7}
Pawar S, Ramchandran K.
\newblock {FFAST: An algorithm for computing an exactly k-Sparse DFT in O(k log
  k) time}.
\newblock IEEE Transactions on Information Theory. 2018;64(1):429--450.
\newblock doi:{10.1109/TIT.2017.2746568}.

\bibitem{bib8}
Pawar S, Ramchandran K.
\newblock {A robust sub-linear time R-FFAST algorithm for computing a sparse
  DFT}. 2015; p. 1--35.

\bibitem{bib9}
Ong F, Heckel R, Ramchandran K.
\newblock {A Fast and Robust Paradigm for Fourier Compressed Sensing Based on
  Coded Sampling}.
\newblock ICASSP, IEEE International Conference on Acoustics, Speech and Signal
  Processing - Proceedings. 2019;2019-May:5117--5121.
\newblock doi:{10.1109/ICASSP.2019.8682063}.

\bibitem{bib10}
Iwen MA.
\newblock {Combinatorial sublinear-time Fourier algorithms}.
\newblock Foundations of Computational Mathematics. 2010;10(3):303--338.
\newblock doi:{10.1007/s10208-009-9057-1}.

\bibitem{bib11}
Iwen MA.
\newblock {Improved approximation guarantees for sublinear-time Fourier
  algorithms}.
\newblock Applied and Computational Harmonic Analysis. 2013;34(1):57--82.
\newblock doi:{10.1016/j.acha.2012.03.007}.

\bibitem{bib12}
Lawlor D, Wang Y, Christlieb A.
\newblock {Adaptive Sub-Linear Time Fourier Algorithms arXiv : 1207 . 6368v1 [
  math . NA ] 26 Jul 2012}. 2012; p. 1--24.

\bibitem{bib13}
Christlieb A, Lawlor D, Wang Y.
\newblock {A multiscale sub-linear time Fourier algorithm for noisy data}.
\newblock Applied and Computational Harmonic Analysis. 2016;40(3):553--574.
\newblock doi:{10.1016/j.acha.2015.04.002}.

\bibitem{bib14}
Merhi S, Zhang R, Iwen MA, Christlieb A.
\newblock {A New Class of Fully Discrete Sparse Fourier Transforms: Faster
  Stable Implementations with Guarantees}.
\newblock Journal of Fourier Analysis and Applications. 2019;25(3):751--784.
\newblock doi:{10.1007/s00041-018-9616-4}.

\bibitem{bib15}
Hassanieh H, Indyk P, Katabi D, Price E.
\newblock {Simple and practical algorithm for sparse fourier transform}.
\newblock Proceedings of the Annual ACM-SIAM Symposium on Discrete Algorithms.
  2012; p. 1183--1194.
\newblock doi:{10.1137/1.9781611973099.93}.

\bibitem{bib16}
Hassanieh H, Indyk P, Katabi D, Price E.
\newblock {Nearly optimal sparse fourier transform}.
\newblock Proceedings of the Annual ACM Symposium on Theory of Computing. 2012;
  p. 563--577.
\newblock doi:{10.1145/2213977.2214029}.

\bibitem{bib17}
Li B, Jiang Z, Chen J.
\newblock {On Performance of Multiscale Sparse Fast Fourier Transform
  Algorithm}. 2020; p. 1--21.

\bibitem{bib18}
Li B, Jiang Z, Chen J.
\newblock {On performance of sparse fast fourier transform algorithms using the
  flat window filter}.
\newblock IEEE Access. 2020;8:79134--79146.
\newblock doi:{10.1109/ACCESS.2020.2989327}.

\bibitem{bib19}
Gilbert AC, Indyk P, Iwen M, Schmidt L.
\newblock {Recent Developments in the Sparse Fourier Transform}.
\newblock IEEE Signal Processing Magazine. 2014; p. 1--21.
\newblock doi:{10.1109/MSP.2014.2329131}.

\bibitem{bib20}
Indyk P, Kapralov M, Price E.
\newblock {(Nearly) sample-optimal sparse fourier transform}.
\newblock Proceedings of the Annual ACM-SIAM Symposium on Discrete Algorithms.
  2014; p. 480--499.
\newblock doi:{10.1137/1.9781611973402.36}.

\bibitem{bib21}
Kapralov M.
\newblock {Sample efficient estimation and recovery in sparse FFT via isolation
  on average}.
\newblock Annual Symposium on Foundations of Computer Science - Proceedings.
  2017;2017-Octob(1):651--662.
\newblock doi:{10.1109/FOCS.2017.66}.

\bibitem{bib22}
Chen GL, Tsai SH, Yang KJ.
\newblock {On Performance of Sparse Fast Fourier Transform and Enhancement
  Algorithm}.
\newblock IEEE Transactions on Signal Processing. 2017;65(21):5716--5729.
\newblock doi:{10.1109/TSP.2017.2740198}.

\bibitem{bib23}
L{\'{o}}pez-Parrado A, {Velasco Medina} J.
\newblock {Efficient Software Implementation of the Nearly Optimal Sparse Fast
  Fourier Transform for the Noisy Case}.
\newblock Ingenier{\'{i}}a y Ciencia. 2015;11(22):73--94.
\newblock doi:{10.17230/ingciencia.11.22.4}.

\bibitem{bib24}
Wang C, Chandrasekaran S, Chapman B.
\newblock {CusFFT: A High-Performance Sparse Fast Fourier Transform Algorithm
  on GPUs}.
\newblock In: Proceedings - 2016 IEEE 30th International Parallel and
  Distributed Processing Symposium, IPDPS 2016. Institute of Electrical and
  Electronics Engineers Inc.; 2016. p. 963--972.

\bibitem{bib25}
Wang C, Chandrasekaran S, Chapman B.
\newblock {CusFFT: A High-Performance Sparse Fast Fourier Transform Algorithm
  on GPUs}.
\newblock In: Proceedings - 2016 IEEE 30th International Parallel and
  Distributed Processing Symposium, IPDPS 2016. Institute of Electrical and
  Electronics Engineers Inc.; 2016. p. 963--972.

\bibitem{bib26}
Abari O, Hamed E, Hassanieh H, Agarwal A, Katabi D, Chandrakasan AP, et~al.
\newblock {A 0.75-million-point fourier-transform chip for frequency-sparse
  signals}.
\newblock In: Digest of Technical Papers - IEEE International Solid-State
  Circuits Conference; 2014.

\bibitem{bib27}
Kapralov M, Velingker A, Zandieh A.
\newblock {Dimension-independent sparse fourier transform}.
\newblock In: Proceedings of the Annual ACM-SIAM Symposium on Discrete
  Algorithms; 2019.

\bibitem{bib28}
Wang S, Patel VM, Petropulu A.
\newblock {Multidimensional Sparse Fourier Transform Based on the Fourier
  Projection-Slice Theorem}.
\newblock IEEE Transactions on Signal Processing. 2019;67(1):54--69.
\newblock doi:{10.1109/TSP.2018.2878546}.

\bibitem{bib29}
Kumar GG, Sahoo SK, Meher PK.
\newblock {50 Years of FFT Algorithms and Applications}. vol.~38.
\newblock Springer US; 2019.
\newblock Available from: \url{https://doi.org/10.1007/s00034-019-01136-8}.

\bibitem{bib30}
Pang C, Liu S, Han Y.
\newblock {High-speed target detection algorithm based on sparse fourier
  transform}.
\newblock IEEE Access. 2018;6:37828--37836.
\newblock doi:{10.1109/ACCESS.2018.2853180}.

\bibitem{bib31}
Plonka G, Wannenwetsch K.
\newblock {A sparse fast Fourier algorithm for real non-negative vectors}.
\newblock Journal of Computational and Applied Mathematics. 2017;321:532--539.
\newblock doi:{10.1016/j.cam.2017.03.019}.

\bibitem{bib32}
Plonka G, Wannenwetsch K.
\newblock {A deterministic sparse FFT algorithm for vectors with small
  support}.
\newblock Numerical Algorithms. 2016;71(4):889--905.
\newblock doi:{10.1007/s11075-015-0028-0}.

\end{thebibliography}
\end{document}